\documentclass[aps,pre,amsfonts,reprint,superscriptaddress]{revtex4-1}
\usepackage{amsmath}
\usepackage{graphicx}
\usepackage{color}

\newcommand{\comment}[1]{}

\begin{document}

\title{Quantum resonant activation} %in modulated spin-boson systems}

\author{Luca Magazz\`u}
\affiliation{Institute of Physics, University of Augsburg, Universit\"atsstrasse 1, D-86135 Augsburg, Germany}
\author{Peter H\"anggi}
\affiliation{Institute of Physics, University of Augsburg, Universit\"atsstrasse 1, D-86135 Augsburg, Germany}
\affiliation{Nanosystems Initiative Munich, Schellingstra{\ss}e 4, D-80799 M\"unchen, Germany}
\affiliation{Department of Physics, National University of Singapore, Singapore 117546}
\affiliation{Centre for Quantum Technologies, National University of Singapore, 3 Science Drive 2, Singapore 117543, Singapore}
\author{Bernardo Spagnolo}
\affiliation{Dipartimento di Fisica e Chimica, Group of Interdisciplinary Theoretical Physics, Universit\`a di Palermo and CNISM,
Unit\`a di Palermo, Viale delle Scienze, Edificio 18, I-90128 Palermo, Italy}
\affiliation{Radiophysics Department, Lobachevsky State University of Nizhny Novgorod, Russia}
\affiliation{Istituto Nazionale di Fisica Nucleare, Sezione di Catania, Italy}
\author{Davide Valenti}
\affiliation{Dipartimento di Fisica e Chimica, Group of Interdisciplinary Theoretical Physics, Universit\`a di Palermo and CNISM,
Unit\`a di Palermo, Viale delle Scienze, Edificio 18, I-90128 Palermo, Italy}

\date{\today}

\begin{abstract}
Quantum resonant activation is investigated for the archetype  setup of  an externally  driven  two-state (spin-boson) system  subjected to strong dissipation  by means of both  analytical and extensive numerical calculations. The phenomenon of resonant activation emerges in the presence of either randomly  fluctuating or deterministic periodically varying driving fields. Addressing the incoherent regime, a characteristic minimum emerges in  the mean  first passage time to reach an absorbing neighboring state whenever the intrinsic time scale of the modulation matches the characteristic time scale of the system dynamics.
For the case of deterministic periodic driving,  the first passage time probability density function (pdf) displays a complex, multi-peaked behavior, which depends crucially on the details of  initial phase,  frequency, and strength of the driving. As an interesting feature we find that the mean  first passage time enters the resonant activation regime at a critical frequency $\nu^*$  which depends  very weakly on the strength of the driving. Moreover, we provide the relation between the first passage time pdf and the statistics of residence times.
\end{abstract}

\maketitle

\section{Introduction}
\label{intro}
The objective of the escape dynamics out of a metastable state has been thoroughly investigated since the seminal work of Kramers~\cite{Hanggi1990}. A quantity of primary interest to establish the time scale of the escape dynamics of a classical Brownian particle in the presence of a potential barrier is the mean first passage time (MFPT); i.e. the average time it takes for a particle driven by noise to reach a target position beyond an intervening  barrier top~\cite{Hanggi1990,Goel1974,Hanggi1983,Talkner1995,Guerin2016}. The topic of evaluating the MFPT in the presence of external modulations of either stochastic or also deterministic nature has ample applications, among others, in neuronal models which are characterized by a time-varying voltage threshold; e.g. see Refs.~\cite{Schindler2004,Ushakov2011}.\\
\indent A minimum occurring in the MFPT versus increasing frequency scale of the modulation is  known in the literature as resonant activation. The phenomenon may emerge when the  time scale of the barrier modulation matches  the  characteristic time scale of the escape dynamics. The phenomenon  was  originally  predicted  for a confining potential composed of a stylized piecewise linear, fluctuating barrier in Ref.~\citep{Doering1992}. Soon after, the objective for the corresponding reaction rate dynamics in presence of  general modulations of a metastable potential
landscape has been investigated with a pioneering work  in Ref.~\cite{Pechukas1994}; cf. also the  surveys on escape over fluctuating barriers~ \cite{Madureira1995,*Madureira1995erratum,Reimann1997},  as well as related studies on  nonequilibrium, dichotomic noise-driven average life times \cite{Mantegna2000,Dubkov2004}. A closely related phenomenon occurs if  periodically varying modulations are acting: resonant activation emerges then due to the interplay  between the  nonstationary, deterministic barrier modulation and thermal ambient noise driven activated escape~\cite{Schmitt2006}.\\
\indent Resonant activation constitutes therefore an  archetypical feature for escape under deterministic modulations or fluctuations of a potential barrier. The general features of the MFPT  as a function of the characteristic modulation frequency scale are a saturation to a maximal value for very slow modulations, where the highest barrier configuration dominates the barrier passage, followed by a decreasing behavior towards an intermediate nonadiabatic minimum -- the resonant activation minimum --, and then by an increase towards a limiting high-frequency behavior, as determined by the corresponding averaged potential configuration~\cite{Doering1992,Pechukas1994,Reimann1997}.\\
\indent With the present work we investigate the phenomenon of resonant activation for the archetype of the widely studied quantum dissipative two-state system (TSS)~\cite{Leggett1987,*Leggett1987erratum, Weiss2008},  here driven by dichotomous noise and/or by a deterministic coherent field. This setup allows for a detailed investigation of the regime in which the barrier is not thermally surmounted  but rather crossed by dissipative quantum tunneling connecting the left- and  right-well states of a double-well exhibiting a lowest energy doublet  of energy separation $\hbar\Delta_0$. Throughout the following we assume  that this lowest doublet is well separated from higher lying quantum energy levels; put differently, the presence of applied modulation is assumed  not to excite appreciably higher lying quantum energies.\\
\indent Modulations of the tunneling amplitude by means of an applied dichotomous noise allow for an exact averaging over the noise realizations. This in turn yields a generalized non-Markovian master equation that has been invoked previously  to investigate the effects of correlated noise  in electron transfer and tight-binding models~\cite{Goychuk1995,Goychuk1998,*Goychuk1998erratum,Goychuk2005}.\\
\indent Given a quantum context,  the concept of the MFPT generally presents, however, a subtle issue~\cite{Muga2000}  originating from the fact that position and momentum knowingly cannot be sharply defined simultaneously. Nevertheless, this issue is overcome when investigating the incoherent tunneling regime. In the latter limit, being realized by coupling the TSS  strongly to an environment, one is able to describe the tunneling dynamics  in terms of a  generally  non-Markovian  quantum master equation for the left/right state  probabilities with  well-defined quantum transition rates. The resulting treatment then mimics (in its Markovian limit)  a classical discrete process~\cite{Goel1974,Talkner2005}. Nevertheless, the dynamics is governed by the quantum  tunneling mechanism; the effect of the environment being a  renormalization of the bare tunneling  amplitude. In such a situation a sensible statement of the problem of an absorbing boundary state in presence of generally time-dependent driving is feasible~\cite{Halliwell1999}.\\
\indent An idealized description of the measurement setting that implements the absorbing state is found in Ref.~\cite{Halliwell1999}, where the detector couples to the particle only in a given region of space. Experimentally, the strongly dissipative regime of incoherent tunneling  in quantum TSS is attained, for example, in superconducting  qubits~\cite{Leggett1987,*Leggett1987erratum,Weiss2008,Caldeira1981,Han1991,Forn-Diaz2017}. Moreover, in recent experiments~\cite{Wagner2017}, a time-resolved detection of tunneling charges is performed using highly controllable devices, such as quantum dots, which are subject to noise and allow for stochastic or deterministic modulations of the tunneling rates.\\
\indent The strategy adopted in the present study is based on the use of a master equation approach to the time-dependent escape dynamics~\cite{Reimann1999}. In this  master equation for the modulated TSS we introduce the appropriate boundary conditions for reflection and absorption.  Particularly, we consider the case with the particle initially prepared in the reflecting left-well state  and set as an absorbing state the neighboring  right-well state. As a result, we end up with an equation for the so-adjusted  decay of the survival  probability $P_L(t)$ in the left metastable state in terms of explicit time-dependent rates. The negative of the rate of change of $P_L(t)$ then defines the  first passage time density -- correctly obeying the boundary conditions at all times --  whose first moment yields  the searched for MFPT.\\
\indent  In the case of dichotomous noise driving, it becomes possible to solve analytically the resulting master equation for the noise-averaged left-well population, at least in the case in which no time-periodic modulations are  present. The latter situations does require a first passage analysis with arbitrary  time-dependent transition rates entering the corresponding quantum master equation; a situation that  can  be treated by numerical means only.\\
\indent Our approach extends the amply studied case of the dissipative   quantum dynamics for a TSS to  the situation with a quantum resonant activated escape regime. Our findings display similar  features as those observed for the mean residence time statistics  occurring in  modulated classical double-well systems where the barrier is thermally surmounted~\cite{Schmitt2006}.  The obtained results therefore corroborate our expectation  that the general phenomenon of resonant activation occurs  likewise in the deep incoherent quantum  regime.  Note also that this used setup  distinctly differs from the stochastic Schr\"odinger equation approach~\cite{Ankerhold2000} and, alike, the approximated semiclassical approach~\cite{Ghosh2005}. Indeed, the study  of quantum resonant activation for a TSS involves a dependence on a wide set of parameters;   it involves, besides the various driving parameters, also the strength and type of quantum dissipation, the temperature, and a suitably chosen  dissipative high-frequency cutoff~\cite{Leggett1987,*Leggett1987erratum,Weiss2008}. This latter value is system-specific as it depends on the  type of physics addressed with such a quantum  TSS~\cite{Goychuk2005}.\\
\section{Driven Quantum Dissipative Two-State dynamics}
\label{model}
As a model of driven dissipative quantum dynamics confined between two metastable  wells, we consider the archetype spin-boson model~\cite{Leggett1987,*Leggett1987erratum} in which a quantum TSS ($S$) is coupled to a  heat bath ($B$) made up of independent bosonic modes of frequencies $\{\omega_i\}$. The coupling to the bath occurs  via  a  scaled position operator which, in the localized basis  $\{|R\rangle,|L\rangle\}$ of a truncated double-well system, is represented by $\sigma_z=|R\rangle\langle R|-|L\rangle\langle L|$.
The total Hamiltonian reads~\cite{Grifoni1998}
\begin{eqnarray}\label{H}
H(t)&=&H_S(t)+H_{SB}+H_B \nonumber\\
&=&-\frac{\hbar}{2}\left[ \Delta(t) \sigma_x+\epsilon(t)\sigma_z\right] \nonumber\\
&-&\frac{\hbar}{2}\sigma_z\sum_i c_i (a_i^{\dag}+a_i)+\sum_i\hbar\omega_i a_i^{\dag}a_i\;,
\end{eqnarray}
where  $\Delta(t)$ denotes  the TSS  tunneling matrix element, modulated around its bare value $\Delta_0$, and $\epsilon(t)$  stands for a modulated  bias energy of vanishing average. Here, $\sigma_x=|R\rangle\langle L|+|L\rangle\langle R|$. In the following sections we consider both deterministic and  stochastic modulations of the tunneling amplitude $\Delta(t)$ and a periodically  driven bias of the form $\epsilon(t)=A_{\epsilon}\cos(\Omega_{\epsilon} t+\phi)$, wherein  $\phi$ denotes an initial phase offset.\\
\indent The bosonic environment, with creation and annihilation operators $a^{\dag}_i$ and $a_i$,  interacts with the TSS system via the set of coupling constants $\{c_i\}$. This system-bath interaction  is  fully characterized by the  spectral density function  $G(\omega)$, whose continuum limit we assume to be of Ohmic form~\cite{Weiss2008}; i.e.,
\begin{eqnarray}\label{G}
 G(\omega)= 2\alpha \omega e^{-\omega/\omega_c},
\end{eqnarray}
 where $\alpha$ characterizes the dimensionless, dissipative coupling strength and $\omega_c$ marks the  suitably chosen high-frequency cutoff.
\subsection{Non-Markovian quantum master equation}
Assuming a factorized initial preparation, with a total density operator of the form $\rho^{\text{tot}}(0)=\rho_{\text{S}}(0)\otimes\rho_{\text{B}}$ (the  bath being initially in the thermal state at temperature $T$), the exact dynamics of the TSS can be cast into the form of a generalized master equation (GME) for the population difference $P(t):= \langle \sigma_z\rangle_t=P_R(t)-P_L(t)$.  Here,  the population $P_{j}(t)=\langle j|\rho_{\text{S}}(t)|j\rangle$   is the probability to find the system in the localized state $j$ ($j=R,L$). The resulting non-Markovian GME assumes  the  form~\cite{Weiss2008,Grifoni1998,Grifoni1996},
\begin{eqnarray}\label{GME}
\dot{P}(t)=\int_{0}^tdt'\;\left[ K^{a}(t,t')-K^{s}(t,t')P(t')\right] \;,
\end{eqnarray}
being formally valid for any coupling and temperature regime, spectral density function, and time dependence of the modulation.
Within the non-interacting blip approximation (NIBA), which is  valid for  strong coupling and not too low temperatures,  these kernels $K^{a/s}$ take on  the explicit expressions~\cite{Grifoni1998}:
\begin{eqnarray}\label{K}
K^{s}(t,t')&=&\Delta(t)\Delta(t') e^{-Q'(t-t')}\cos[Q''(t-t')]\cos[\zeta(t,t')] \nonumber\\
K^{a}(t,t')&=&\Delta(t)\Delta(t') e^{-Q'(t-t')}\sin[Q''(t-t')]\sin[\zeta(t,t')] \nonumber,\\
&\quad&
\end{eqnarray}
where the function $\zeta$ is defined by
\begin{eqnarray}\label{zeta}
\zeta(t,t')=\int_{t'}^t dt''\;\epsilon(t'')\;.
\end{eqnarray}
The kernels $K^{s}(t,t')$ and $K^{a}(t,t')$ in Eq.~(\ref{K}) are  symmetric and antisymmetric, respectively,  under the change $\epsilon(t)\rightarrow-\epsilon(t)$. This implies that, in the static  unbiased case, $K^{a}(t,t') = 0$.\\
\indent The functions $Q'(t)$ and $Q''(t)$ in Eq.~(\ref{K}) denote the real and imaginary part of the thermal bath correlation function, respectively~\cite{Weiss2008}. For the chosen Ohmic spectral density in Eq.~(\ref{G}) 
and in the so-called scaling limit ($k_BT\ll \hbar\omega_c$), $Q(t)$ reads~\cite{Thorwart2001}
\begin{eqnarray}\label{Q}
Q(t)&=&2\alpha\ln\left[\left(1+\omega_{c}^{2} t^{2}\right)^{\frac{1}{2}}\frac{\sinh(\kappa t)}{\kappa t}\right]\nonumber\\
&&+i2\alpha\arctan(\omega_{c}t),
\end{eqnarray}
where $\kappa=\pi k_{B}T/\hbar$.
\subsection{Quantum master equation in the incoherent regime}
In the incoherent tunneling regime, occurring at finite temperatures and {\it strong} coupling (i.e., $\alpha>0.5$ for the symmetric TSS)~\cite{Weiss2008pp358-359}, the nondriven dynamics of the population difference is well approximated by the Markovian limit to Eq.~(\ref{GME}) with time-independent transition rates. This is so because the memory time of the kernels in Eq.~(\ref{K}) constitutes the smallest time scale. In the driven case, using the definition of $P(t)$ and the conservation of total probability, i.e.,  $P_R(t)+P_L(t)=1$, a master equation for the individual probabilities with time-dependent forward ($+$) and backward ($-$) rates is derived. This master equation,  valid for general modulations of $\Delta(t)$ and $\epsilon(t)$~\cite{Grifoni1998,Goychuk1999,Goychuk2006}, reads
\begin{eqnarray}\label{GMEPLR}
\dot{P}_L(t)&=&W^-(t)P_R(t)-W^+(t)P_L(t) \nonumber\\
\dot{P}_R(t)&=&W^+(t)P_L(t)-W^-(t)P_R(t),
\end{eqnarray}
where
\begin{eqnarray}\label{W-pm}
W^{\pm}(t)&=&\frac{\Delta(t)}{2}\int_0^{\infty}d\tau\; \Delta(t-\tau) e^{-Q'(\tau)} \nonumber\\
&&\qquad\times\cos[Q''(\tau)\mp\zeta(t,t-\tau)]
\end{eqnarray}
are the quantum transition rates from the left to the right well (forward) and \emph{vice versa} (backward). In this Markovian limit, the rates generally vary in time, but are independent of the population themselves. These transition  rates incorporate implicitly both the quantum dissipation and the shape of the double-well potential and depend as well  only locally on the externally applied modulation.\\
\indent We note that setting the upper integration limit to infinity in Eq.~(\ref{W-pm}) constitutes a further approximation within this type of Markovian limit. However, carrying out the integration up to the latest acting physical time $\tau=t$, yields an improvement which allows us to analyze the short time behavior of the pdf. The resulting improved Markovian quantum transition rates read
\begin{eqnarray}\label{W-pm-exact}
\mathcal{W}^{\pm}(t)&=&\frac{\Delta(t)}{2}\int_0^{t}d\tau\; \Delta(t-\tau) e^{-Q'(\tau)} \nonumber\\
&&\qquad\times\cos[Q''(\tau)\mp\zeta(t,t-\tau)]\;.
\end{eqnarray}
The upper integration limit  at $\infty$ in Eq.~(\ref{W-pm}) yields time-independent rates in the absence of deterministic driving.  This modification is of relevance  for the regime of very  short passage times only. In practice, using $\infty$ as the upper integration limit produces indistinguishable numerical results away from the very short time regime. In the main panels of Figs.~\ref{fig2},~\ref{fig6},~\ref{fig10}, and~\ref{fig12}, shown in the result section $IV$ below, we have consistently used the improved rate expression~(\ref{W-pm-exact}). We confirmed numerically that noticeable differences occur only in the short-time regime shown in the insets.\\
\indent In the case of time-periodic modulations $\Delta(t)$ or $\epsilon(t)$, this difference in the transient behavior corresponds to the fact that the rates in Eq.~(\ref{W-pm}) are strict periodic functions of $t$, while those in Eq.~(\ref{W-pm-exact}) acquire the same periodic behavior only at times larger than the memory time of the kernels.\\
\section{First passage time dynamics}
\label{sec-MFPT}
\subsection{Time-dependent boundary conditions}
\label{boundary-cond}
\begin{figure}[ht!]
\begin{center}
\includegraphics[width=0.3\textwidth,angle=0]{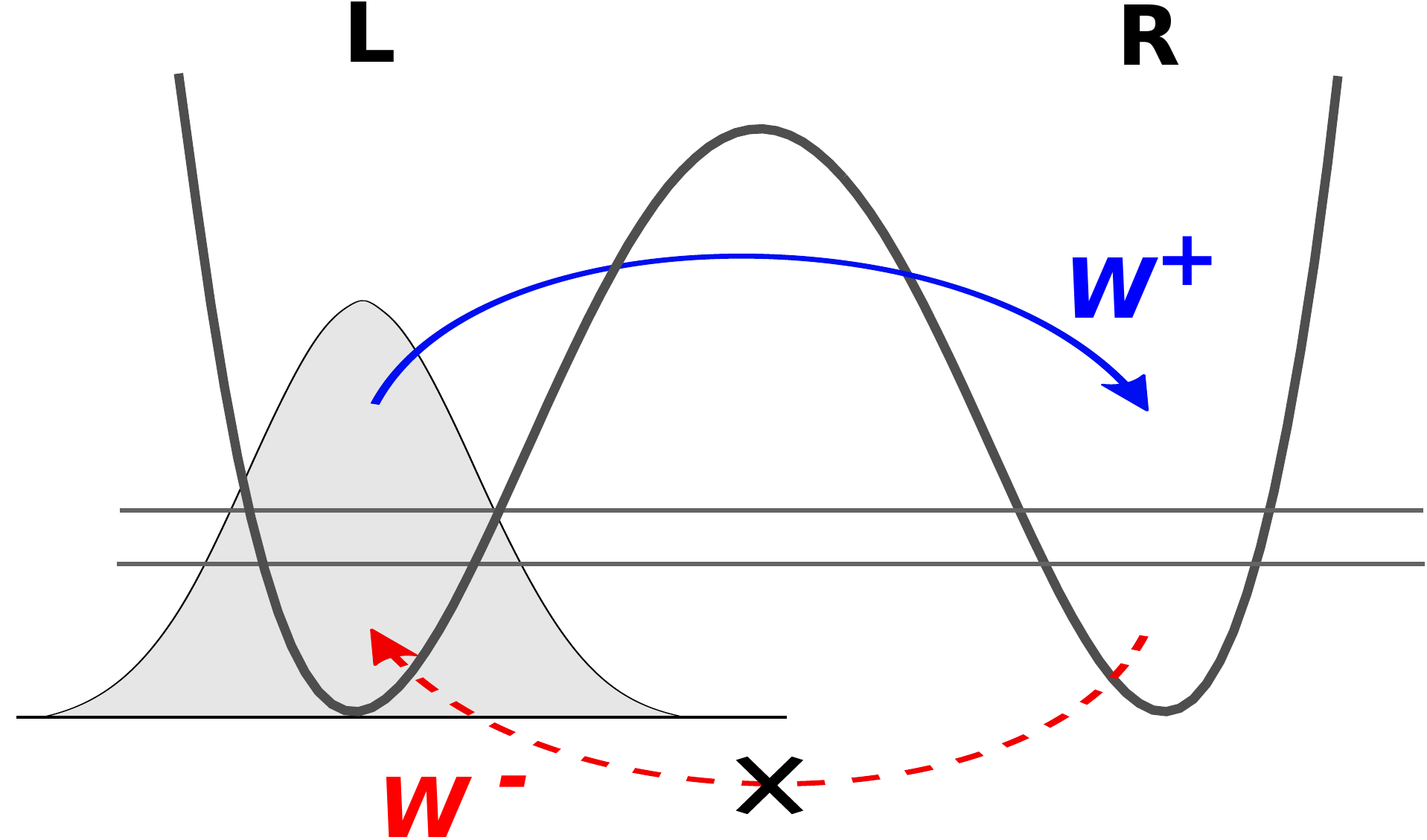}
\caption{\footnotesize{(Color online) Particle initially in the left metastable well of a double-well potential in the two state system approximation. The right-well state is an absorbing state.}}
\label{fig1}
\end{center}
\end{figure}
The quantum master equation, Eq.~(\ref{GMEPLR}), can be understood as  describing a discrete stochastic process, randomly switching between two \emph{reflecting} states; meaning that the rates to go  leftward at the left state and rightward at the opposite, right-placed state are both vanishing~\cite{Goel1974pp13-14}.
To perform a first passage time analysis, we consider the situation in which the particle is initially prepared at time $t = 0$ in the left quantum state ($L$). We next calculate the passage time statistics to become detected (absorbed) at the right  state ($R$) while the left state is kept reflecting~\cite{Goel1974pp13-14}. This requirement is implemented upon introducing an absorbing boundary conditions at the state $R$ and reflecting boundary condition  at the state $L$.  Given these two  generally  time-dependent ``birth and death'' quantum transition rates, this amounts to setting for all times $t\geq 0$~\cite{Goel1974pp13-14}
\begin{eqnarray}\label{boundary-cond}
W^-(t)=0\qquad\text{and}\qquad W^+(t)>0
\end{eqnarray}
in Eq.~(\ref{GMEPLR}); see Fig.~\ref{fig1}. Moreover, given the initial condition that $P_L(0)=1$, the left well population $P_L(t)$  must be interpreted as the \emph{conditional survival probability} $P(L;t|L;0)$. This conditional  survival probability in the left state, with $R$ absorbing state, is thus  governed by
\begin{eqnarray}\label{eqPL}
\dot{P}_L(t)=-W^+(t)P_L(t),
\end{eqnarray}
with initial condition $P_L(0)=1$ and forward rate $W^+(t)$ detailed with  Eq.~(\ref{W-pm}). Note that this conditional probability distinctly differs from the ones governed by Eq.~(\ref{GMEPLR}).  \\
\indent Just alike in a classical situation~\cite{Talkner2003,Schindler2004,Schindler2005}, the negative rate of change of this so-tailored  conditional passage time probability to find the particle still in state $L$  yields  the  first-passage time (FPT) probability density function (pdf) which is given by
\begin{eqnarray}\label{g}
g(t)=-\dot{P}_L(t)\;,
\end{eqnarray}
with $\dot{P}_L(t)$ determined from  Eq.~(\ref{eqPL}).
With  positive-valued forward rates and starting out at $P_L(t=0)=1$ we have, with  absorption occurring  at state $R$, that  $P_L(t=\infty)=0$. The FPT pdf $g(t)$ in Eq. ~(\ref{g})  satisfies $g(t)\geq 0$ and is properly normalized, i.e.,  $\int_0^{\infty} dt\; g(t)=1$. Moreover, by using the improved expression~(\ref{W-pm-exact}) for the rate, $g(t)$ then starts out at $g(t=0)= 0$.\\
\indent The  MFPT to the state $R$ of the TSS can be obtained in the commonly known  way ~\cite{Hanggi1990,Hanggi1983},  namely as the first moment $t_1$ of the FPT pdf $g(t)$ in Eq.~(\ref{g}); reading
\begin{eqnarray}\label{MFPT}
t_1=\int_0^{\infty}dt \; t g(t)\;.
\end{eqnarray}
\indent In the following we focus on this first moment, as it constitutes the quantity of interest for our analysis of the resonant activation. However, the knowledge of $g(t)$, given by Eq.~(\ref{g}) upon solving Eq.~(\ref{eqPL}), allows for the calculation of higher moments of the FPT pdf. These quantities provide additional information on the passage time statistics, possibly of relevance for experimental realizations. For example, fluctuations around the MFPT, quantified by the second moment, provide a  measure of the number of detections needed to collect a reliable statistics for the FPT analysis.\\
\indent The FPT pdf also determines the so-termed residence time and interspike pdfs, which generally are more readily available in experiments, e.g., in the context of stochastic resonance phenomena \cite{Gammaitoni1998}, and involve suitable averages over the FPT pdf \cite{Talkner2003,Schindler2005,Schindler2004,Talkner2005}. The residence time pdf is explicitly evaluated in Sec.~\ref{driven-Delta-zero-bias},  in the context of applying the theory to the case of  periodic modulations of the tunneling element; see Fig.~\ref{fig9} below. 
\subsection{Time-periodic modulation}
\label{sec-periodical-driving}
Up to here, the theory has been  general in regard to the choice for the shape of the temporal  modulation. Here and in the following sections we specify the various specific forms of modulations used in evaluating both the FPT pdf $g(t)$ and its first  mean, the  MFPT  $t_1$. \\
\indent  We start with the case where one of the two parameters of the TSS (either the bare tunneling matrix element $\Delta_0$ or the bias)  is  periodically modulated in time, while the other is held fixed. To be specific, consider the following two forms of  periodically  driven  settings:
\begin{eqnarray}\label{}
\text{i)}\quad \Delta(t)&=&\Delta_0+A_d\cos(\Omega_d t +\phi)\nonumber\\
 \epsilon(t)&=&0\nonumber\\
\text{ii)}\quad \Delta(t)&=&\Delta_0 \nonumber\\
\epsilon(t)&=&A_{\epsilon}\cos(\Omega_{\epsilon} t +\phi)\;.
\end{eqnarray}
For a vanishing amplitude of the driving on the tunneling matrix element, i.e., $A_d=0$ in i), and as well for the bias, i.e., $A_{\epsilon}=0$ in ii), the static case  with $P_L(t)=\exp(-W^+ t)$ is recovered, wherein $W^+=\Delta_0^2/2\int_0^{\infty}d\tau\exp[Q'(\tau)]\cos[Q''(\tau)]$.\\
\indent For both the driving settings, the FPT pdf   depends explicitly on the initial driving phase $\phi$. Consequently, the MFPT  is evaluated as an average over a uniform distribution of this phase, yielding
\begin{eqnarray}\label{g-driving-only}
g(t)=\frac{1}{2\pi}\int_0^{2\pi}d\phi\; g(t;\phi)\;,
\end{eqnarray}
with $g(t;\phi)=-\dot{P}_L(t;\phi)$, and rhs evaluated by Eq.~(\ref{eqPL}) in terms of the phase-dependent quantum transition rate $W^+(t)$.
\subsection{Driving with a combination of dichotomous noise and deterministic periodic driving}
\label{noise-av-ME}
Next, consider a situation in which the system is driven with a deterministic modulation of the bias; i.e.,
\begin{eqnarray}
\label{periodical-epsilon}
\epsilon(t)=A_{\epsilon}\cos(\Omega_{\epsilon} t +\phi)
\end{eqnarray}
and the tunneling amplitude is driven by stationary, exponentially correlated  dichotomous  noise (also known as telegraphic noise) of vanishing average around its bare value $\Delta_0$. Explicitly, we set
\begin{eqnarray}\label{Delta_noise}
\Delta(t)=\Delta_0+\Delta\eta(t),
\end{eqnarray}
where \cite{Hanggi1995} $\eta(t)=(-1)^{n(t)}$ with $n(t)$ a Poissonian counting process with parameter $\nu$, yielding that $\eta^2(t)=1$.  Here, the  amplitude $\Delta$  is a two-state random variable $u$ which is evenly distributed, i.e., $\rho(u)= 0.5 [\delta(u+\Delta) + \delta(u-\Delta)]$, thus having a vanishing average, while  the Poisson parameter $\nu$ determines the noise correlation of the  two-state dichotomous process $\xi(t)= \Delta\eta(t)$, i.e.,
\begin{eqnarray}\label{}
\langle\xi(t)\xi(t')\rangle_{\eta}&=&\Delta^2 e^{-\nu|t-t'|},
\end{eqnarray}
where the subscript $\eta$ stands for average over the noise realizations.\\
\indent In the extreme limit $\nu\rightarrow\infty$ and $\Delta^2 \rightarrow \infty$ this two-state  noise approaches   white Gaussian noise of vanishing mean~\cite{Hanggi1995,Gardiner2004}. Keeping  the noise amplitude fixed, however,  the intensity of this noise vanishes identically with $\nu \rightarrow \infty$, as can be seen by writing the noise correlation function as   $\langle\xi(t)\xi(0)\rangle=(2\Delta^2/\nu) [\nu \exp(-\nu|t|)/2]$, where the term inside the square brackets approaches a Dirac delta-function whereas its strength (i.e. the prefactor) vanishes.
This accounts for the behavior observed in Sec.~\ref{fluctuating-delta} below, where the limit $\nu \rightarrow \infty$  of dichotomous fluctuations indeed coincides with the MFPT for the noiseless case.\\
\indent Dichotomous noise allows for an exact averaging over the noise realizations of the dynamics given by Eq.~(\ref{eqPL}). As detailed in Appendix~\ref{derivation}, the noise-averaged population $\langle P_L(t)\rangle_{\eta}$ is obtained by solving the set of equations in which $\langle P_L(t)\rangle_{\eta}$ is coupled to the correlation expression $y(t)\equiv \langle\eta(t)P_L(t)\rangle_{\eta}$. The rate of change for $\langle P_L(t)\rangle_{\eta}$ is then given by
\begin{eqnarray}\label{eqPLfluct}
\langle \dot{P}_L(t) \rangle_{\eta}&=&-W^+_0(t) \langle P_L(t) \rangle_{\eta}-W^+_1(t)y(t) \nonumber\\
\dot{y}(t)&=&-W^+_1(t) \langle P_L (t)\rangle_{\eta}-\left[ W^+_0(t)+\nu\right] y(t).\qquad
\end{eqnarray}
The value of $y(t)$ at $t=0$ gives the initial correlation between the position of the particle and the state of the noise $\eta(t=0)=\pm1$.  In what follows we assume uncorrelated initial condition, i.e.,  Eq.~(\ref{eqPLfluct}) is solved with initial conditions $\langle P_L(t)\rangle_{\eta}=1$ and $y(t=0)=0$. The rates appearing in Eq.~(\ref{eqPLfluct}) are obtained as
\begin{eqnarray}\label{W-pm-i}
W^+_i(t)&=&\frac{1}{2}\int_0^{\infty}d\tau\; S_{i}(\tau) e^{-Q'(\tau)}\cos[Q''(\tau)\mp\zeta(t,t-\tau)] \nonumber\\
&&
\end{eqnarray}
with $i=0,1$  and  where
\begin{eqnarray}\label{S}
S_0(t)&=&\Delta_0^2+\Delta^2e^{-\nu t}\nonumber\\
S_1(t)&=&\Delta_0\Delta\left( 1+e^{-\nu t}\right).
\end{eqnarray}
Because the noise amplitude appears as the prefactor in the function $S_1(t)$, it follows readily that, for vanishing noise amplitude $\Delta=0$, $W^+_1(t) = 0$. The averaged probabilities then decouple from $y(t)$. In this case the first line  of Eq.~(\ref{eqPLfluct}) reduces to an equation formally identical to Eq.~(\ref{eqPL}) with $\Delta(t)=\Delta_0$. Also note that, as stated  before for the rates~(\ref{W-pm}), here too the time-dependent rates must  be properly defined with the upper integration limit set to $t$, i.e.,
\begin{eqnarray}\label{W-pm-i-exact}
\mathcal{W}^+_i(t)&=&\frac{1}{2}\int_0^{t}d\tau\; S_{i}(\tau) e^{-Q'(\tau)}\cos[Q''(\tau)-\zeta(t,t-\tau)]\nonumber\\
&&
\end{eqnarray}
and again i=0,1. \\
\indent The MFPT  is calculated by using the FPT pdf averaged over the two-state noise realizations and also over the initial phase of the deterministic driving, yielding
\begin{eqnarray}\label{g-driving-noise}
g(t) &=&\frac{1}{2\pi}\int_0^{2\pi}d\phi\; \langle g[t;\phi;\eta(t)]\rangle_{\eta},
\end{eqnarray}
where $\langle g[t;\phi;\eta(t)]\rangle_{\eta}=-\langle \dot{P}_L(t;\phi)\rangle_{\eta}$, with $\langle P_L(t;\phi)\rangle_{\eta}$ given by Eq.~(\ref{eqPLfluct}) with phase-dependent rates.\\
\indent Before showing the results of our analysis of the resonant activation, it is important to note that the theory developed above, aside from the specific expressions of the spin-boson NIBA rates, is completely general and applies to generic systems where the rates of incoherent tunneling are subject to periodic modulations and/or to dichotomous noise.
\section{Results}
\label{results}
This section reports the findings for  the  resonant activation occurring in an incoherent spin-boson system with modulated tunneling matrix element and/or oscillating bias. Specifically, we consider for the  tunneling matrix element $\Delta(t)$ separate modulations: Either an unbiased two-state noise $\eta(t)$ or a deterministic periodic driving. Only afterwards we consider a more  general case for which  the tunneling matrix element is fluctuating  while a periodically oscillating field  drives the bias $\epsilon(t)$. Throughout the remaining parts all quantities are scaled in terms of the bare tunneling frequency $\Delta_0$; i.e.,
\begin{itemize}
\item Frequencies $\Omega_d$, $\Omega_{\epsilon}$, and $\omega_c$ and noise switching rate  $\nu$ are in units of $\Delta_0$. Time $t$ is measured in units of $\Delta_0^{-1}$.
\item Temperature T is measured in units of $\hbar\Delta_0/k_B$.
\end{itemize}
As can be deduced from Hamiltonian~(\ref{H}), noise and driving amplitudes are frequencies and are thus given in units of $\Delta_0$. Moreover, in what follows  cutoff frequency and  temperature are held fixed, assuming the values $\omega_c=10$ and $T=0.2$. Finally, with the exception of the results in  Figs.~\ref{fig5} and~\ref{fig8}, the dimensionless coupling strength is always set to the value $\alpha=0.7$.
\subsection{Dichotomously fluctuating tunneling matrix element in absence of a bias energy: Analytical treatment}
\label{fluctuating-delta}
\begin{figure}[ht!]
\begin{center}
\includegraphics[width=0.5\textwidth,angle=0]{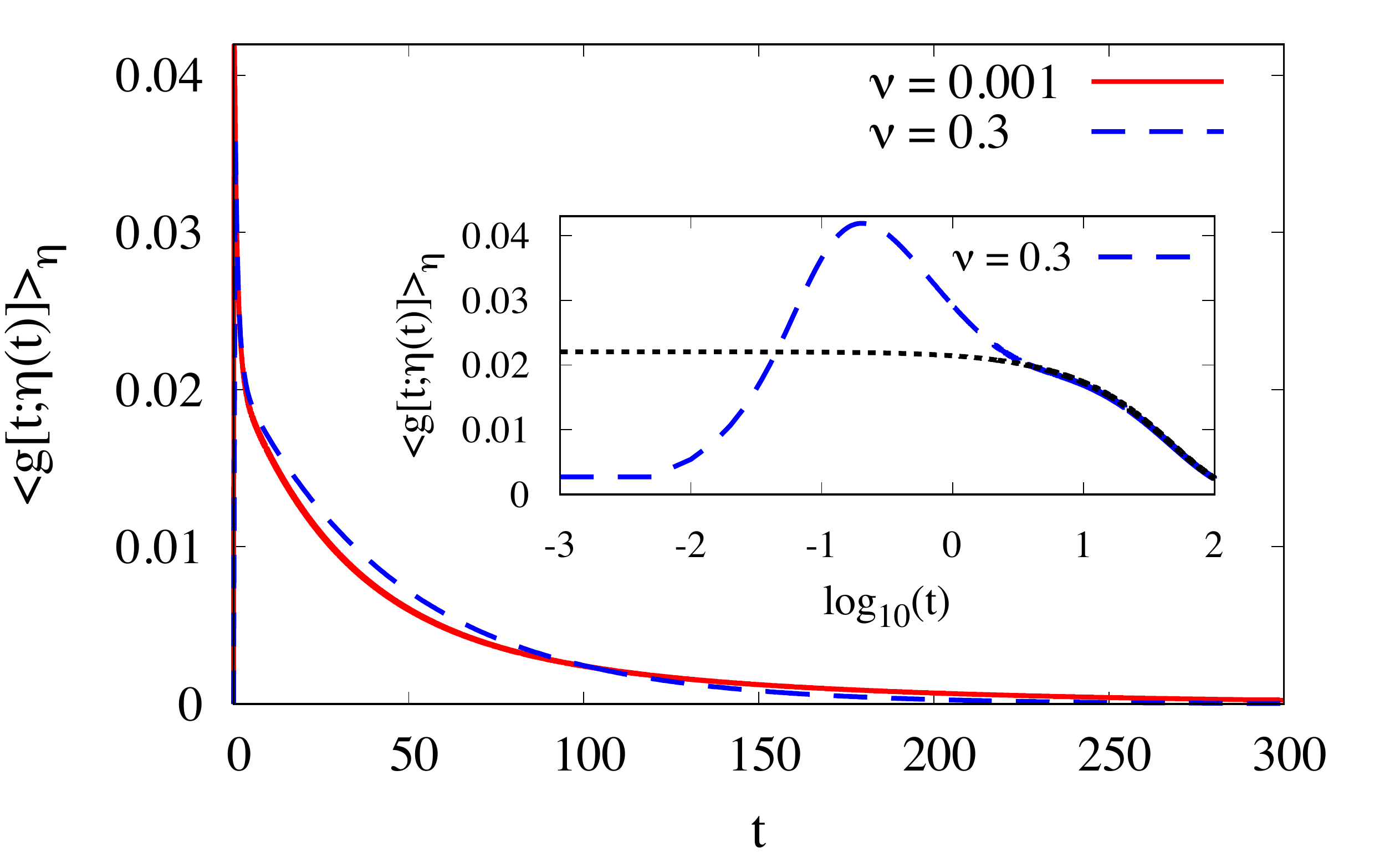}
\caption{\footnotesize{(Color online) First passage time pdf for two-state noise modulating the tunneling matrix element and constant zero bias.  Adiabatic Poisson rate (red solid line) and  nonadiabatic, intermediate noise switching regime  (blue dashed line).  The noise strength is  set to $\Delta=0.3$.
Inset: Close-up of the $\nu=0.3$ curve in $\mathrm\log_{10}$ scale. Dotted line: Same quantity  evaluated using the time-independent transition rate calculated according to Eq.~(\ref{W-pm-i}) with $\zeta(t,t')=0$.
The remaining parameters are $\alpha=0.7$, $T=0.2$, and $\omega_c=10$.}}
\label{fig2}
\end{center}
\end{figure}

\begin{figure}[ht!]
\begin{center}
\includegraphics[width=0.5\textwidth,angle=0]{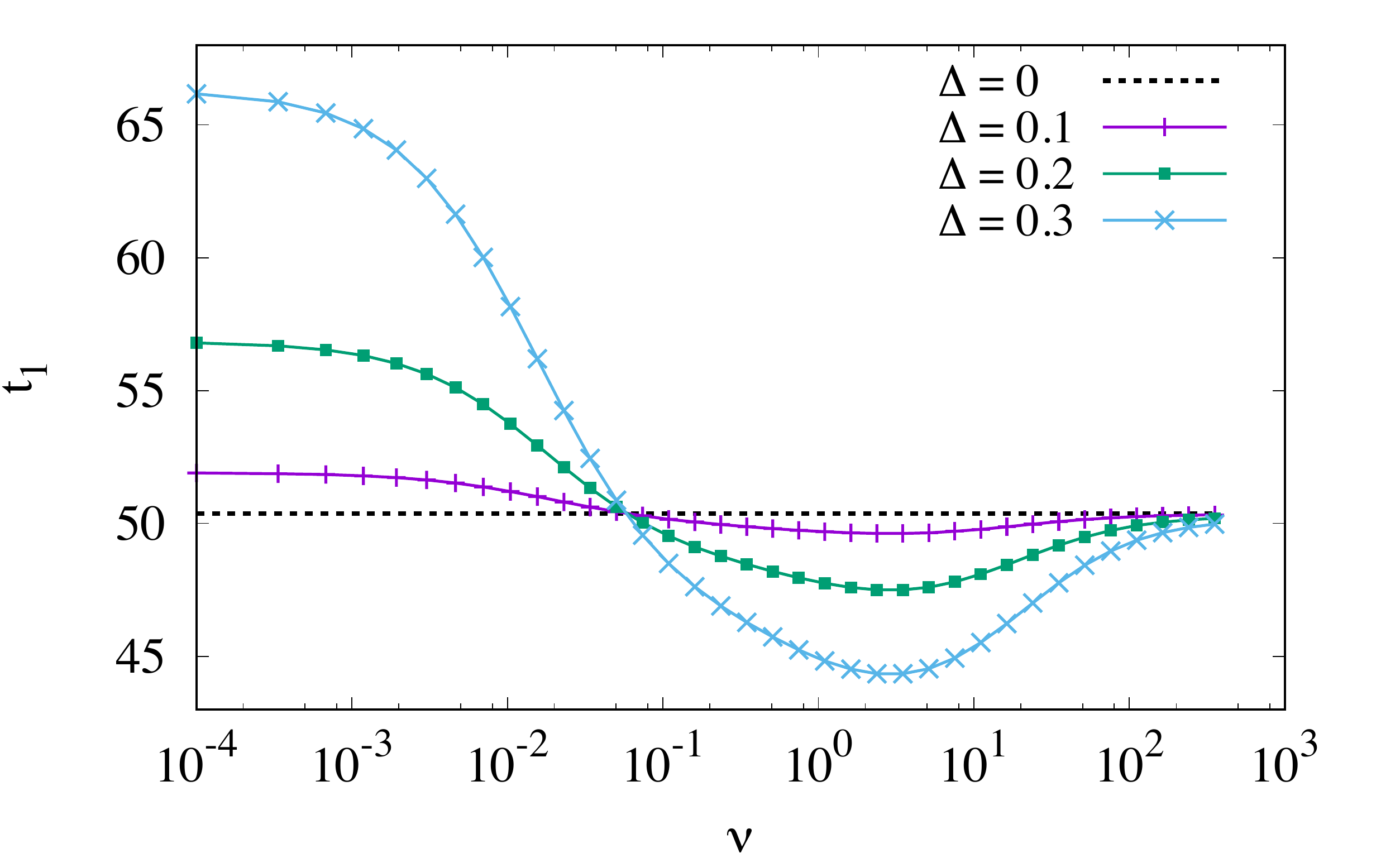}
\caption{\footnotesize{(Color online) Mean first passage time $t_1$ {\it vs.} Poisson rate $\nu$ for two-state fluctuations of the tunneling matrix element with different amplitudes $\Delta$ and constant zero bias, as given analytically by Eq.~(\ref{MFPT-analytical}). The remaining parameters  are as in Fig.~\ref{fig2}.}}
\label{fig3}
\end{center}
\end{figure}

This situation with stationary telegraphic noise modulations can be treated  analytically.  We first address the setting described in Sec.~\ref{noise-av-ME} in this analytically solvable case in which the bias is vanishing, i.e.,  $A_{\epsilon}=0$, and the tunneling matrix element fluctuates around its static value $\Delta_0$, namely
\begin{eqnarray}\label{}
\epsilon(t)&=&0\nonumber\\
\Delta(t)&=&\Delta_0+\Delta\eta(t)\;.
\end{eqnarray}
The noise $\xi(t)=\Delta\eta(t)$ denotes the   Markovian two-state noise of vanishing average, as  discussed in Sec.~\ref{noise-av-ME}.
 This setting corresponds to a TSS version of the classical Brownian particle in a piecewise linear fluctuating potential considered by Doering and Gadoua in Ref.~\cite{Doering1992}, but here the dynamics is governed by quantum  tunneling transition rates rather than by classical (Arrhenius-like)  over-barrier escape  rates.\\
\indent In this case Eq.~(\ref{eqPLfluct}) for the noise-averaged population reads explicitly
\begin{eqnarray}\label{ME_noise_Delta}
\langle \dot{P}_L (t)\rangle_{\eta}&=&-W^+_0 \langle P_L (t)\rangle_{\eta}-W^+_1 y(t) \nonumber\\
\dot{y}(t)&=&-W^+_1 \langle P_L (t)\rangle_{\eta}-\left( W^+_0 +\nu\right) y(t),
\end{eqnarray}
where the time-independent transition rates are given by Eq.~(\ref{W-pm-i}) with $\zeta(t,t')=0$. Using Eqs.~(\ref{W-pm-i}) and~(\ref{S}), the rates $W^+_i$ are explicitly given by
\begin{eqnarray}\label{rates2}
W^+_0&=&\Delta_0^2a_0+\Delta^2 a_\nu \nonumber\\
W^+_1&=&\Delta_0\Delta (a_0+a_\nu),
\end{eqnarray}
where
\begin{eqnarray}\label{a}
a_\nu&=&\frac{1}{2}\int_0^{\infty}d\tau\; e^{-\nu\tau-Q'(\tau)}\cos[Q''(\tau)]
\end{eqnarray}
and $a_0 \equiv a_{\nu=0}$.\\
\indent Due to the fact that the transition rates are time-independent, Eq.~(\ref{ME_noise_Delta}) can be solved analytically with the boundary conditions $ \langle P_L (0)\rangle_{\eta}=1$ and $y(0)=0$. The solution of Eq.~(\ref{ME_noise_Delta}) for the noise-averaged  population of the left state is
\begin{subequations}\label{solution}
\begin{eqnarray}
\langle P_L (t)\rangle_{\eta}=C_1e^{-\gamma_1 t}+C_2 e ^{-\gamma_2 t}\:,
\end{eqnarray}
where
\begin{eqnarray}
C_{1/2}=\frac{d\pm\nu}{2d}\quad\text{and}\quad
\gamma_{1/2}=\frac{2 W^+_0+\nu \mp d}{2},
\end{eqnarray}
 and further
\begin{eqnarray}
 d=\sqrt{\nu^2+4(W^+_1)^2}\;.
\end{eqnarray}
\end{subequations}
Note that, for vanishing noise amplitude, i.e.,  $\Delta=0$, the rate $W^+_1$ vanishes identically. In this latter case  Eq.~(\ref{solution}) renders a strictly single-exponential decay with $\langle P_L (t)\rangle_{\eta}=\exp(-\Delta_0^2 a_0 t)$.\\
\\
\indent The FPT pdf, i.e.,
\begin{eqnarray}\label{g-noise}
g(t)&=&\langle g[t;\eta(t)]\rangle_{\eta}\nonumber\\
&=&-\langle \dot{P}_L(t)\rangle_{\eta}
\end{eqnarray}
assumes the form of a bi-exponential decay, as it follows from taking the time derivative of the solution in Eq.~(\ref{solution}). In Fig.~\ref{fig2} we depict $g(t)$  for two values of the Poisson parameter  $\nu$. For the lower, adiabatic rate (red solid line) g(t) overrides the corresponding nonadiabatic curve (blue dashed line) at long times, having a larger tail. This gives rise to a MFPT $t_1$ whose value, in the adiabatic case, exceeds the one assumed in the intermediate (nonadiabatic) regime, as can be seen in  Fig.~\ref{fig3}. In Fig.~\ref{fig2} we  plot the numerically evaluated pdf  $g(t)$, using the time-dependent rates given by Eq.~(\ref{W-pm-i-exact}) with $\zeta(t,t')=0$. The inset in  Fig.~\ref{fig2} shows that  evaluations for $g(t)$,  using either the time-dependent expression or the time-independent form~(\ref{W-pm-i}),  yield  results that differ at very short times only, but otherwise become indistinguishable! For this reason the MFPT evaluated alternatively with those time-independent transition rates provides an excellent approximation. This feature also  holds true alike for the FPT pdfs calculated in subsequent sections.  \\
The MFPT $t_1$ as a function of the Poisson rate $\nu$ can be calculated analytically by using the  solution~(\ref{solution}) in  Eq.~(\ref{g-noise}) and the definition for $t_1$ given in Eq.~(\ref{MFPT}). We find that
\begin{eqnarray}\label{MFPT-analytical}
t_1(\nu)&=&\frac{C_1(\nu)}{\gamma_1(\nu)}+\frac{C_2(\nu)}{\gamma_2(\nu)}\nonumber\\
&=&\frac{W^+_0(\nu)+\nu}{(W^+_0(\nu))^2+\nu W^+_0(\nu)-(W^+_1(\nu))^2}\;,
\end{eqnarray}
where the dependence on the Poisson  parameter $\nu$ is made explicit.
From this analytic result three important limits can be investigated using Eqs.~(\ref{rates2}) and~(\ref{a}). First, the static case is recovered upon setting $\Delta=0$, yielding
\begin{eqnarray}\label{t1-static}
t_{1,\Delta=0}=(\Delta_0^2a_0)^{-1}.
\end{eqnarray}
This same value is assumed by the MFPT in the limit $\nu\rightarrow \infty$; i.e.,
\begin{eqnarray}\label{t1-high-rate}
\lim_{\nu \to \infty}t_1(\nu)&=&(\Delta_0^2a_0)^{-1} \nonumber\\
&=&t_{1,\Delta=0}.
\end{eqnarray}
Finally, in  the adiabatic  limit $\nu\rightarrow 0$, the MFPT emerges as
\begin{eqnarray}\label{t1-low-rate}
\lim_{\nu \to 0}t_1(\nu)=\frac{1}{a_0}\frac{\Delta_0^2+\Delta^2}{(\Delta_0^2-\Delta^2)^2}.
\end{eqnarray}
\indent In Fig.~\ref{fig3} the MFPT $t_1$, evaluated according to Eq.~(\ref{MFPT-analytical}), is depicted as a function of the Poisson  rate $\nu$ for different values of the noise amplitude $\Delta$.  The curves display adiabatic, low switching rate maxima whose values, for different values of $\Delta$, approach  the analytical limit~(\ref{t1-low-rate}). The resonantly activated regime occurs at intermediate noise switching time scales. As described by Eqs.~(\ref{t1-static}) and~(\ref{t1-high-rate}), at large  noise switching rates $\nu$ the MFPT converges to the results for the \emph{average} configuration which, in our case, coincides with the unmodulated, static case. These general features are shared with the predictions obtained in Refs.~\cite{Doering1992,Pechukas1994,Bier1993}  using a classical Brownian motion escape dynamics.\\
\indent Fig.~\ref{fig3} depicts yet another intriguing feature: The different curves seemingly cross exactly the horizontal line (static case) at a switching  rate  which surprisingly depends very weakly on the noise amplitude $\Delta$. Interestingly, a similar behavior has also been observed numerically in Ref.~\cite{Fiasconaro2011} for  classical Brownian particle dwelling a piecewise linear fluctuating barrier and in experiments~\cite{Miyamoto2010}.\\
\indent Analytical evaluations of the MFPT for a wider range of amplitudes indicate that this \emph{crossing point} for entering the resonant activation regime in the nonadiabatic regime   is in fact  mathematically {\it not} exact; see the  filled circles  in Fig.~\ref{fig4} below. This  near-exact nonadiabatic crossing frequency $\nu^*$, at which  the MFPT crosses the static value, cf.  Fig.~\ref{fig3},  is determined by the solution to the transcendental equation
\begin{eqnarray}\label{equation}
\frac{C_1(\nu^{*})}{\gamma_1(\nu^{*})}+\frac{C_2(\nu^{*})}{\gamma_2(\nu^{*})}=\frac{1}{\Delta_0^2a_0}\;.
\end{eqnarray}
The above relation results from equating the analytical expression of the MFPT  in Eq.~(\ref{MFPT-analytical}) with the static value Eq.~(\ref{t1-static}) as given by the dotted line in Fig.~\ref{fig3}.\\
\indent The relation in Eq.~(\ref{equation}) can be solved approximately in analytical terms  by assuming that the rates $W^+_i$ are nearly  \emph{independent} of the Poisson rate $\nu$, when restricted to a narrow regime  around the {\it a posteriori} chosen numerical value $\nu\sim 0.06$. Put differently, we substitute  $a_\nu$ with the value
$\tilde{a}:=a_{\nu=0.06}$ in Eq.~(\ref{rates2}) (with the present choice of parameters,  numerical values for the coefficients are $a_0=1.985\times10^{-2}$ and $\tilde{a}=2.102\times10^{-2}$). The solution of Eq.~(\ref{equation}) for the crossing point then reads
\begin{eqnarray}\label{crossing}
\nu^{*}\simeq\Delta_0^2\left(\frac{a_0^2}{\tilde{a}}+a_0+\tilde{a}\right)-\Delta^2 \tilde{a}\;.
\end{eqnarray}
This shows  that the leading contribution to $\nu^*$ is quadratic in the amplitude  $\Delta$ so that, for $\Delta<1$, the crossing point depends only weakly  on this noise amplitude. The analytic  crossing rate $\nu^*$ obtained  from Eq.~(\ref{crossing})  as a function of $\Delta$ is shown in Fig.~\ref{fig4} as the solid (violet) line. Note the excellent quantitative agreement between this approximate evaluation of the crossing rate $\nu^*$ and the numerically precise evaluation at selected noise amplitudes marked by the filled circles.\\
\begin{figure}[ht!]
\begin{center}
\includegraphics[width=0.5\textwidth,angle=0]{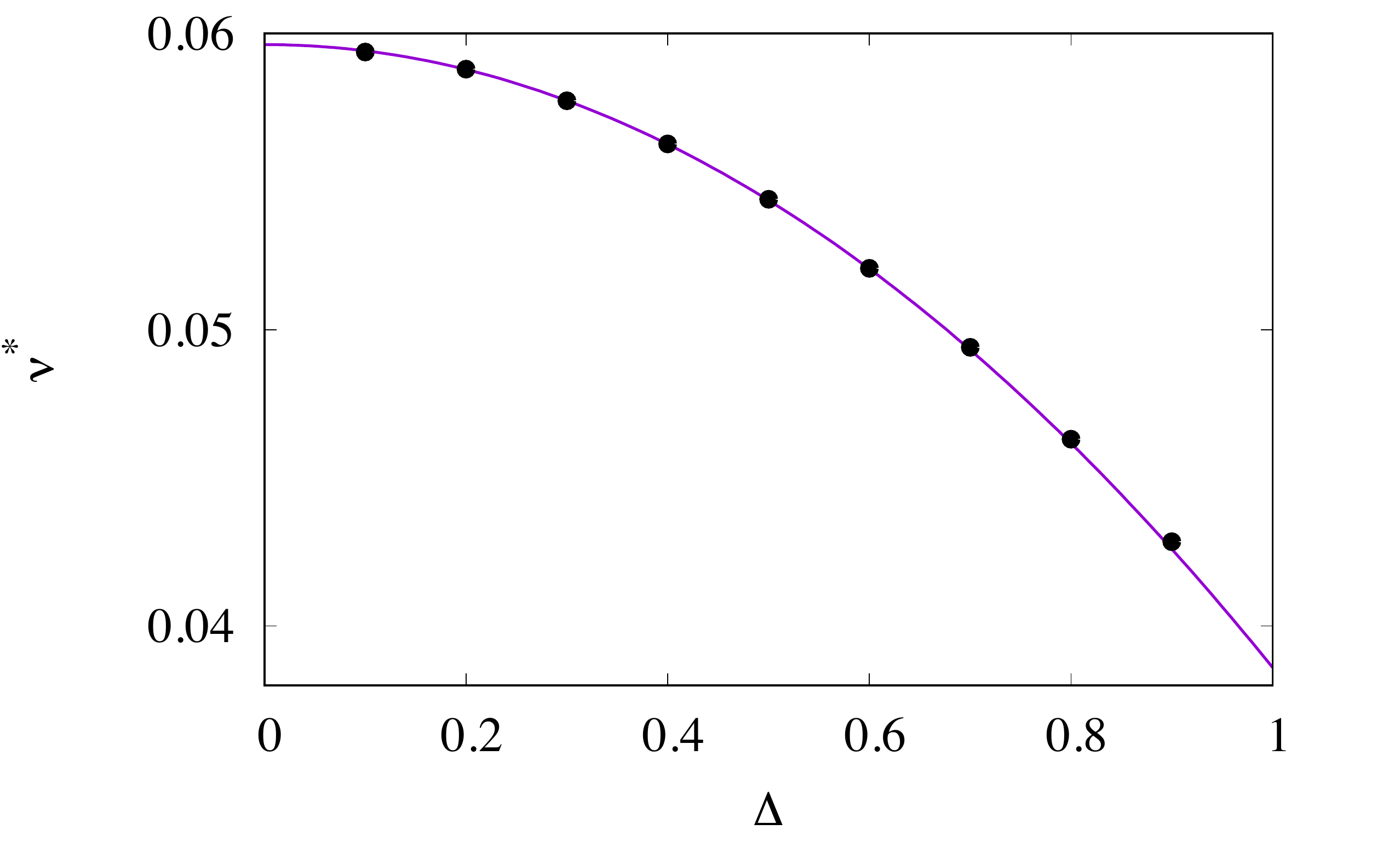}
\caption{\footnotesize{(Color online) Near-crossing point behavior for the MFPT. Approximate solution, Eq.~(\ref{crossing}), for the crossing Poisson rate $\nu^*$ versus the noise amplitude $\Delta$  (solid line) for dichotomous fluctuations of the tunneling matrix element and with $\epsilon(t)=0$. Filled circles: - Numerical precise values from relation Eq.~(\ref{equation}), evaluated at selected noise amplitudes $\Delta$. The remaining parameters are as in  Fig.~\ref{fig2}.}}
\label{fig4}
\end{center}
\end{figure}
\begin{figure}[ht!]
\begin{center}
\includegraphics[width=0.5\textwidth,angle=0]{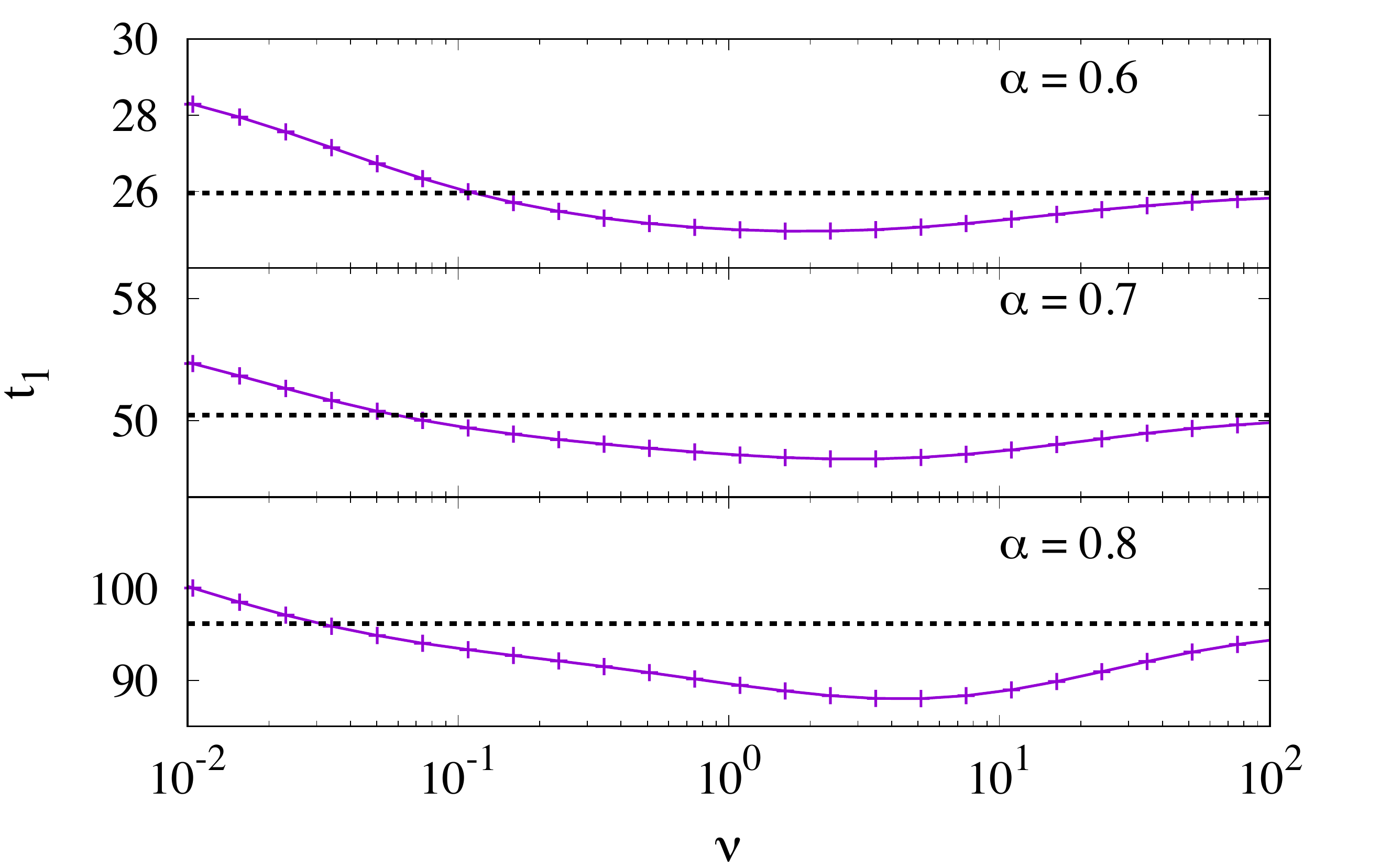}
\caption{\footnotesize{(Color online)(solid lines) Mean first passage time $t_1$ {\it vs.} the noise switching rate $\nu$ for dichotomous fluctuations of the tunneling matrix element with noise amplitude  $\Delta=0.2$  and $\epsilon(t)=0$. The panels depict the results for  different dissipation strengths $\alpha$. The dotted lines indicate the static cases with $\Delta=0$ with a bare static value $\Delta_0=1$ (this value is fixed at $1$  within our choice made for dimensionless units).  The curves in the central panel coincide with evaluations in Fig.~\ref{fig3}. Other parameters are as in Fig.~\ref{fig2}.}}
\label{fig5}
\end{center}
\end{figure}
\indent To provide a deeper insight for the interplay between the characteristic time scale of the dynamics, essentially dictated by $\alpha$, and that of the noise dynamics, encoded in the Poisson parameter $\nu$, we show in Fig.~\ref{fig5} a comparison among the MFPT results {\it vs.} $\nu$ at different values of the dissipation strength $\alpha$. The crossing frequency $\nu^*$ assumes a lower value (corresponding to larger noise correlation time) for stronger bath coupling, where the bare tunneling amplitude becomes  dissipation-renormalized towards a lower value \cite{Weiss2008}, meaning that the tunneling passage to the rightward well occurs on a larger time scale. We also observe that the regime of noise switching  rates for  resonant activation spans a wider regime with increasing  dissipation strength. See also Fig.~\ref{fig8} below, where the same  features are obtained with deterministic, periodic modulation of the tunneling amplitude. \\
\indent Summarizing, upon increasing the rate of the dichotomous noise modulating  the bare tunneling element $\Delta_0$, the MFPT goes across the three distinctive regimes  depicted in Fig.~\ref{fig3}: It saturates to a maximal value in the limit of adiabatically slow  modulations and then monotonically decreases towards the resonant activation minimum at intermediate values of the noise rate. This minimum is in turn followed by a monotonic increase towards an intermediate limiting value at high noise rates.  The latter coincides with the value of the MFPT in the noiseless  case.\\
\indent This very general behavior can be accounted for with the following argument, which is along the lines of  that put forward in Ref~\cite{Pechukas1994} for a classical process with fluctuating barriers. In the adiabatic regime the modulation is slower than the relaxation in the slower static configuration. Thus the latter dominates the FPT density. In the opposite limit  of fast modulations, the system is subject to an average configuration yielding a lower value of the MFPT. Finally, when the modulation is slow enough that an  \emph{instantaneous} rate can be individuated on the driving time scale but fast with respect to the relaxation dynamics, then the dynamics results from the average rate over the system's configurations. Now, this average rate is larger than the rate of the average configuration, given the dependence of the rate on the value of the tunneling element set by Eq.~(\ref{W-pm}), and results in the resonant activation minimum of the MFPT. Note also that, in the present incoherent regime, an increase of the coupling causes a slower relaxation dynamics~\cite{Weiss2008}. This, in turn, makes the above-discussed condition for the onset of the resonant activation regime valid at lower noise rates (the noise is fast with respect to the relaxation dynamics already at low Poisson rates), consistently with what is observed in Fig.~\ref{fig5}.
\subsection{Periodically varying modulations: Numerical treatment}
\label{sec-pure_driving}
\subsubsection{Periodically driven  tunneling matrix element}
\label{driven-Delta-zero-bias}
Here we consider a situation in which the tunneling matrix  element  is subjected to a periodically varying, deterministic driving of the form
\begin{eqnarray}\label{driven-Delta}
\epsilon(t)&=&0\nonumber\\
\Delta(t)&=&\Delta_0+A_d\cos(\Omega_d t+\phi)\;.
\end{eqnarray}
\begin{figure}[ht!]
\begin{center}
\includegraphics[width=0.5\textwidth,angle=0]{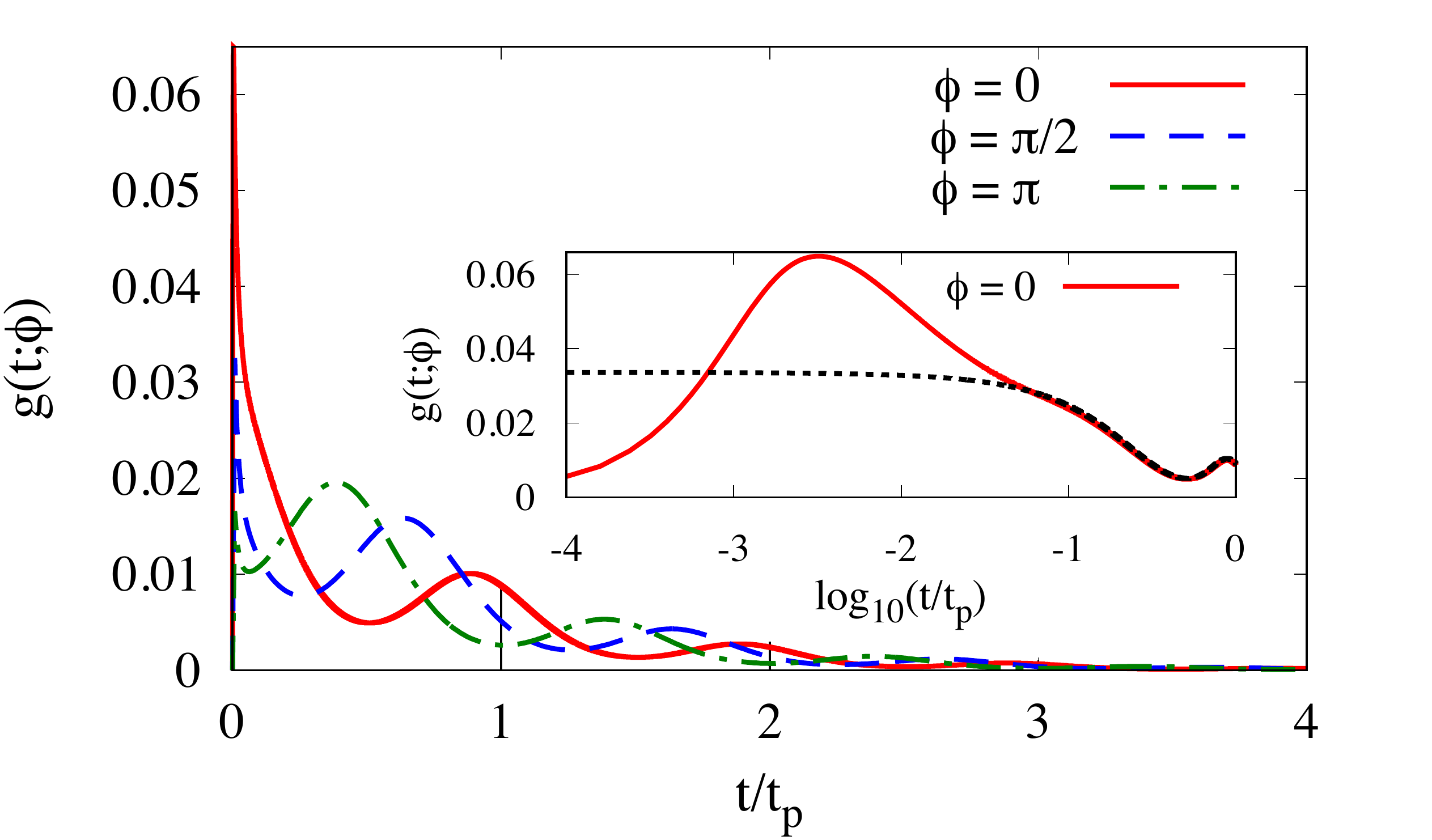}
\caption{\footnotesize{(Color online)  First passage time pdf for a periodically driven  tunneling matrix element, i.e., $\Delta(t)=\Delta_0+A_d\cos(\Omega_d t+\phi)$, with amplitude $A_d=0.3$, period $t_p=2\pi/\Omega_d$, where $\Omega_d=0.1$, for three values of the initial driving phase $\phi$ and with $\epsilon(t)=0$.  Inset: $\phi=0$ curve up to one period $t_p$ in $\mathrm \log_{10}$ scale. Dotted line: Same quantity evaluated using the time-dependent rate calculated according to Eq.~(\ref{W-pm}). The remaining parameters are as in  Fig.~\ref{fig2}.}}
\label{fig6}
\end{center}
\end{figure}
\indent In Fig.~\ref{fig6}  the FPT pdf $g(t;\phi)=-\dot{P}_L(t;\phi)$,  with $P_L(t;\phi)$ given by Eq.~(\ref{eqPL}) and a phase-dependent rate $W^+(t)$, is considered for three values of the phase $\phi$; see Eq.~(\ref{driven-Delta}). The presence of periodic driving causes a modulation on the FPT pdf similar in spirit  to the   FPT pdf obtained for a periodically driven  leaky integrate-and-fire model for neural spiking; -- there, the FPT pdf peaks (for an initial driving phase $\phi=\pi/2$) seemingly tend to synchronize with the driving oscillation period  in the adiabatic limit $\Omega_d \rightarrow 0$. ~\cite{Schindler2004,Schindler2005}. Moreover, this oscillating behavior is similar to that observed for the switching time probability in a long Josephson junction~\cite{Valenti2014}.\\
\begin{figure}[ht!]
\begin{center}
\includegraphics[width=0.5\textwidth,angle=0]{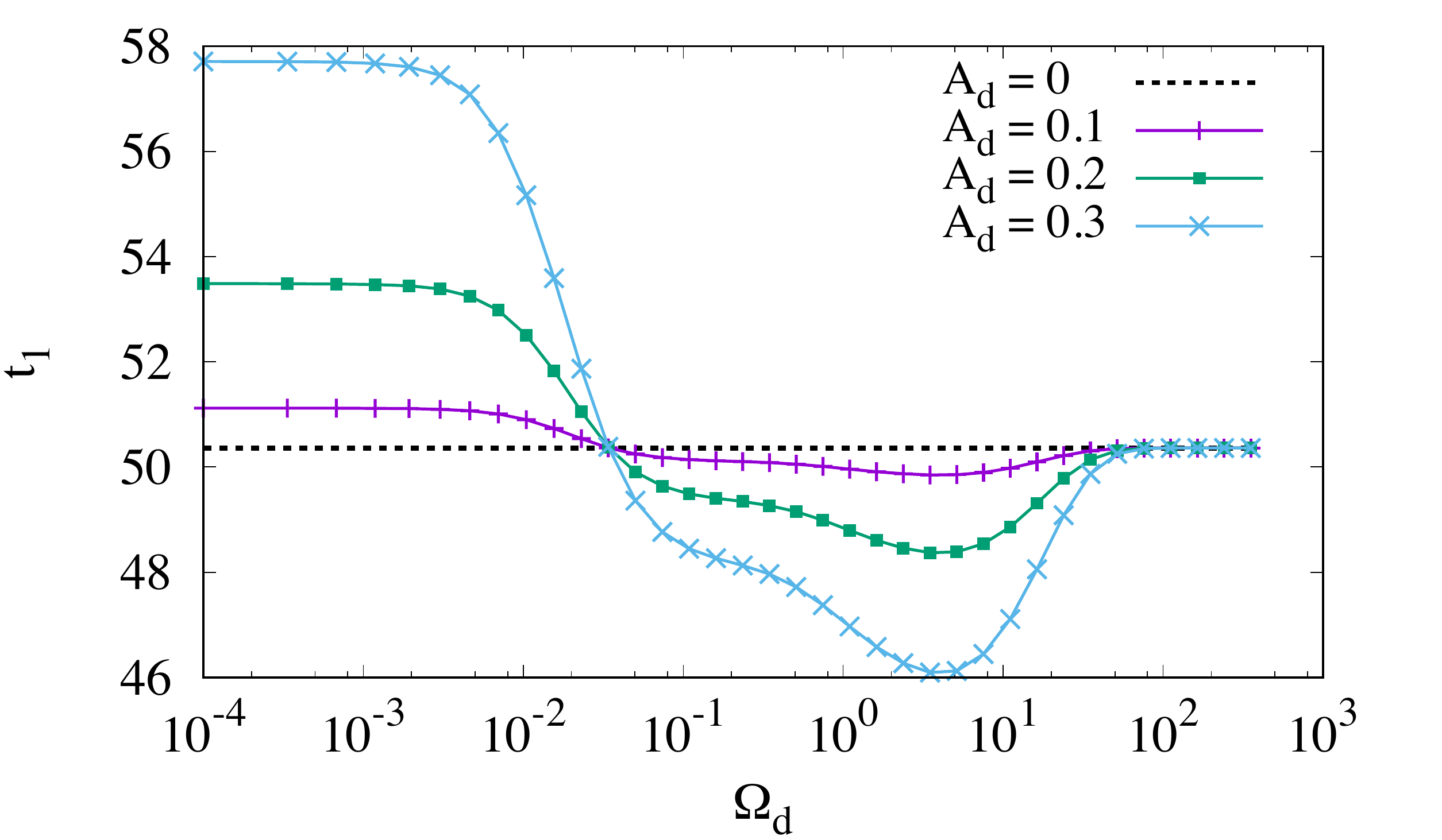}
\caption{\footnotesize{(Color online) Mean first passage time $t_1$ (averaged over the initial phase $\phi$) {\it vs.} angular driving frequency $\Omega_d$ for periodic driving of the tunneling matrix element with  different driving amplitudes $A_d$ and a constant bias  $\epsilon(t)=0$. The remaining parameters are as in  Fig.~\ref{fig2}.}}
\label{fig7}
\end{center}
\end{figure}
\indent The MFPT is obtained by solving Eq.~(\ref{eqPL}) with  time-dependent rate  $W^+(t)$ determined by using Eq.~(\ref{W-pm}) with $\zeta(t,t')=0$ and $\Delta(t)$ from Eq.~(\ref{driven-Delta}).
The MFPT $t_1$ versus the angular driving frequency $\Omega_d$, for different values of the amplitude $A_d$, is shown in Fig.~\ref{fig7}.
For each value of $\Omega_d$ the average over the phase $\phi$ of the driving in Eq.~(\ref{g-driving-only}) is realized by uniformly sampling the interval $[0,2\pi)$ at $40$ intermediate values. The results for the MFPT display essentially the same features as for the noise-driven tunneling matrix element in Fig.~\ref{fig3}; namely, the low frequency saturation to a maximal value at slow driving, the resonant activated regime occurring at intermediate driving frequencies, where $t_1$ underruns  the static value, and the convergence at high frequency to the MFPT value of the average configuration. The latter coincides with the static configuration. Also in this case,  results for a larger driving amplitudes range (not shown) display a nearly exact crossing. This implies that the crossing frequency $\Omega_d^*$, where $t_1$ enters the resonant activation regime (i.e. the crossing with the horizontal line in Fig.~\ref{fig7}), also here depends weakly on the  amplitude $A_d$.\\
\indent In Fig.~\ref{fig8} we compare the obtained MFPTs versus the angular frequency $\Omega_d$ for different values of dissipation strength $\alpha$. The results show the same features already observed in Fig.~\ref{fig5} for the noise-modulated tunneling matrix element.\\
\begin{figure}[ht!]
\begin{center}
\includegraphics[width=0.5\textwidth,angle=0]{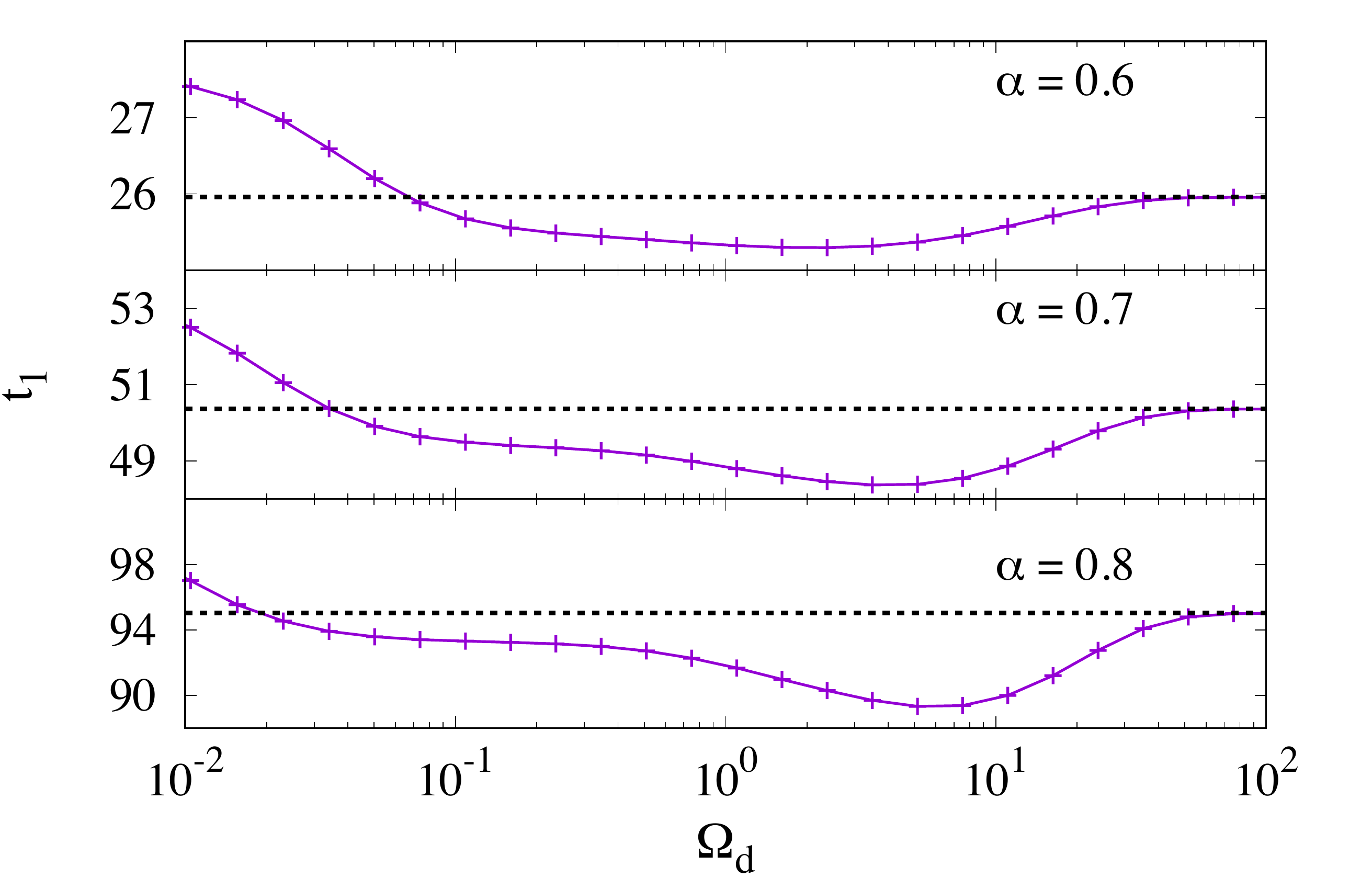}
\caption{\footnotesize{(Color online) Mean first passage time $t_1$ (averaged over $\phi$) {\it vs.} angular frequency $\Omega_d$  of the periodic driving of the tunneling matrix element of amplitude strength $A_d=0.2$ (solid lines) and again $\epsilon(t)=0$. Dotted lines: Static cases ($A_d=0$). Comparison among different dissipation strengths $\alpha$.  Other parameters are  as in Fig.~\ref{fig2}.}}
\label{fig8}
\end{center}
\end{figure}
\indent In concluding this section, we relate the FPT pdf $g(t)$ to a quantity more easily  accessed in actual experiments~\cite{Gammaitoni1998,Schmitt2006}. To this purpose, consider, for the very same driving setup discussed in this section, the following different protocol: Instead of iterating the procedure of preparing the system in state $L$ and resetting the driving phase, after each absorption, imagine that the particle is not absorbed but is left free to reenter the state $L$ after a random time, whose distribution at long times is given by the asymptotic probability of being in state $R$ times the backward rate $W^{-}(t)$. The particle is thus prepared only once in the left well with the phase of the periodic modulation set to, say, $\phi=0$.\\
\indent Then, the quantity of interest, directly accessed in experiments, is  the residence time distribution  $R_L(t)$ (not a pdf) in state $L$ . This distribution is the starting time average over one driving period $t_p$, with normalized asymptotic entrance probability density, of the survival time distribution in the state $L$;  see  Eq. (31) in Ref.~\cite{Talkner2005}. The  quantity $r(t)=-d/dt\; R_L(t)$, i.e.  the pdf of \emph{residence times},  relates directly with the FPT pdf and, for the situation described above, reads
\begin{eqnarray}\label{RTpdf}
r(t)=\frac{\int_0^{t_p}ds\; g(t+s;\phi=0|s)W^-(s)P_R^{\text{as}}(s)}{\int_0^{t_p}ds\; W^-(s)P_R^{\text{as}}(s)},
\end{eqnarray}
where the conditional character of $g(t+s|s)$ -- the particle is transferred into the left state at time $s$ -- is made explicit. $P_R^{\text{as}}(s)$ is the asymptotic value of the  population of state $R$ satisfying Eq.~(\ref{GMEPLR}).
A plot of $r(t)$ is provided in Fig.~\ref{fig9} where a comparison is made with the FPT pdf, both at fixed phase $\phi=0$ and averaged over $\phi$ according to Eq.~(\ref{g-driving-only}).
\begin{figure}[ht!]
\begin{center}
\includegraphics[width=0.5\textwidth,angle=0]{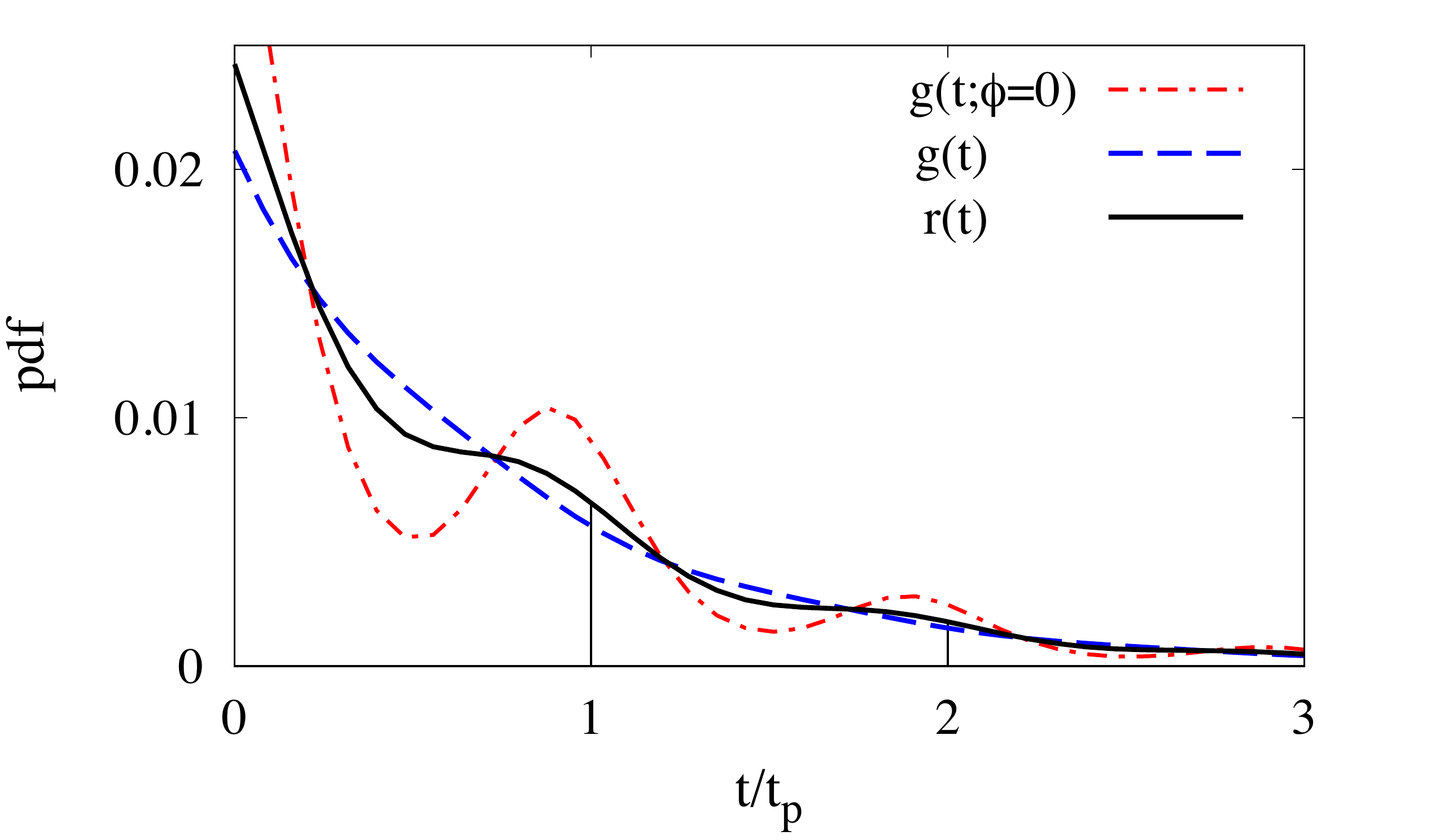}
\caption{\footnotesize{(Color online)  Comparison between the first passage time pdf -- at fixed phase $\phi=0$ (cf. Fig.~\ref{fig6}) and averaged over $\phi$, according to Eq.~(\ref{g-driving-only}) -- and the residence time pdf $r(t)$, as obtained from Eq.~(\ref{RTpdf}). Calculations are performed by using the periodically varying rates in Eq.~(\ref{W-pm}).  Driving setup and parameters are the same as in Fig.~\ref{fig6}.}}
\label{fig9}
\end{center}
\end{figure}
\subsubsection{Periodically oscillating bias and constant tunneling matrix element}
\label{driven-bias}
As a second configuration with purely deterministic modulation we consider the case where the tunneling matrix element is held constant, $\Delta(t)=\Delta_0$, while a periodic driving modulates the  bias $\epsilon(t)$  according to
\begin{eqnarray}\label{epsilon-t}
\epsilon(t)=A_{\epsilon}\cos(\Omega_{\epsilon} t+\phi).
\end{eqnarray}
The population of the left state satisfies formally the same equation as for the periodically driven tunneling matrix element, Eq.~(\ref{eqPL}), with forward transition rate $W^+(t)$ given by Eq.~(\ref{W-pm}) and fixed tunneling  amplitude $\Delta(t)=\Delta_0$.\\
\begin{figure}[ht!]
\begin{center}
\includegraphics[width=0.5\textwidth,angle=0]{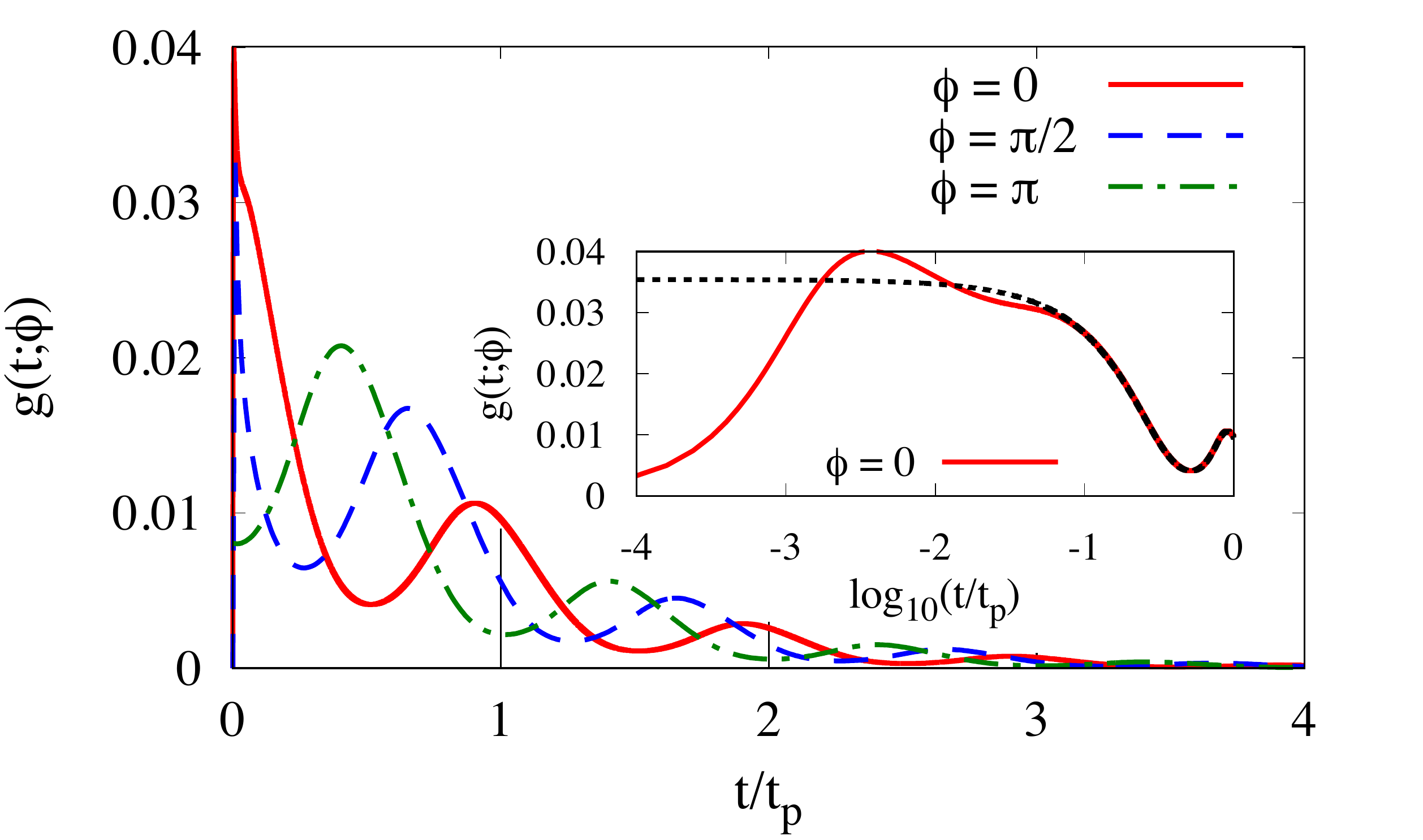}
\caption{\footnotesize{(Color online) First passage time pdf for a periodically  driven bias with amplitude $A_{\epsilon}=0.3$ and period $t_p=2\pi/\Omega_{\epsilon}$ with  $\Omega_{\epsilon}=0.1$.  The three values of the initial driving phase $\phi$ are as in Fig.~\ref{fig6}. Inset: $\phi=0$ curve up to one driving period in $\mathrm \log_{10}$ scale. Dotted line: Same quantity evaluated using the rates calculated according to Eq.~(\ref{W-pm}). Other parameters are as in Fig.~\ref{fig2}.}}
\label{fig10}
\end{center}
\end{figure}
\indent In Fig.~\ref{fig10} the FPT pdf $g(t;\phi)=-\dot{P}_L(t;\phi)$ is depicted for three values of the initial driving phase $\phi$. Also in this case,  as for the setting with periodically driven  tunneling matrix element (cf. Fig.~\ref{fig6}), the FPT pdf displays multiple peaks whose position depends on the fixed phase $\phi$.\\
\indent Results for the MFPT $t_1$ versus the angular  frequency $\Omega_{\epsilon}$, for different values of the driving amplitude $A_{\epsilon}$, are shown in Fig.~\ref{fig11}. As in the previous subsection, also in this case the average over $\phi$ prescribed by Eq.~(\ref{g-driving-only}) is performed  by uniformly sampling the interval $[0,2\pi)$ at $40$ intermediate values.\\
\begin{figure}[ht!]
\begin{center}
\includegraphics[width=0.5\textwidth,angle=0]{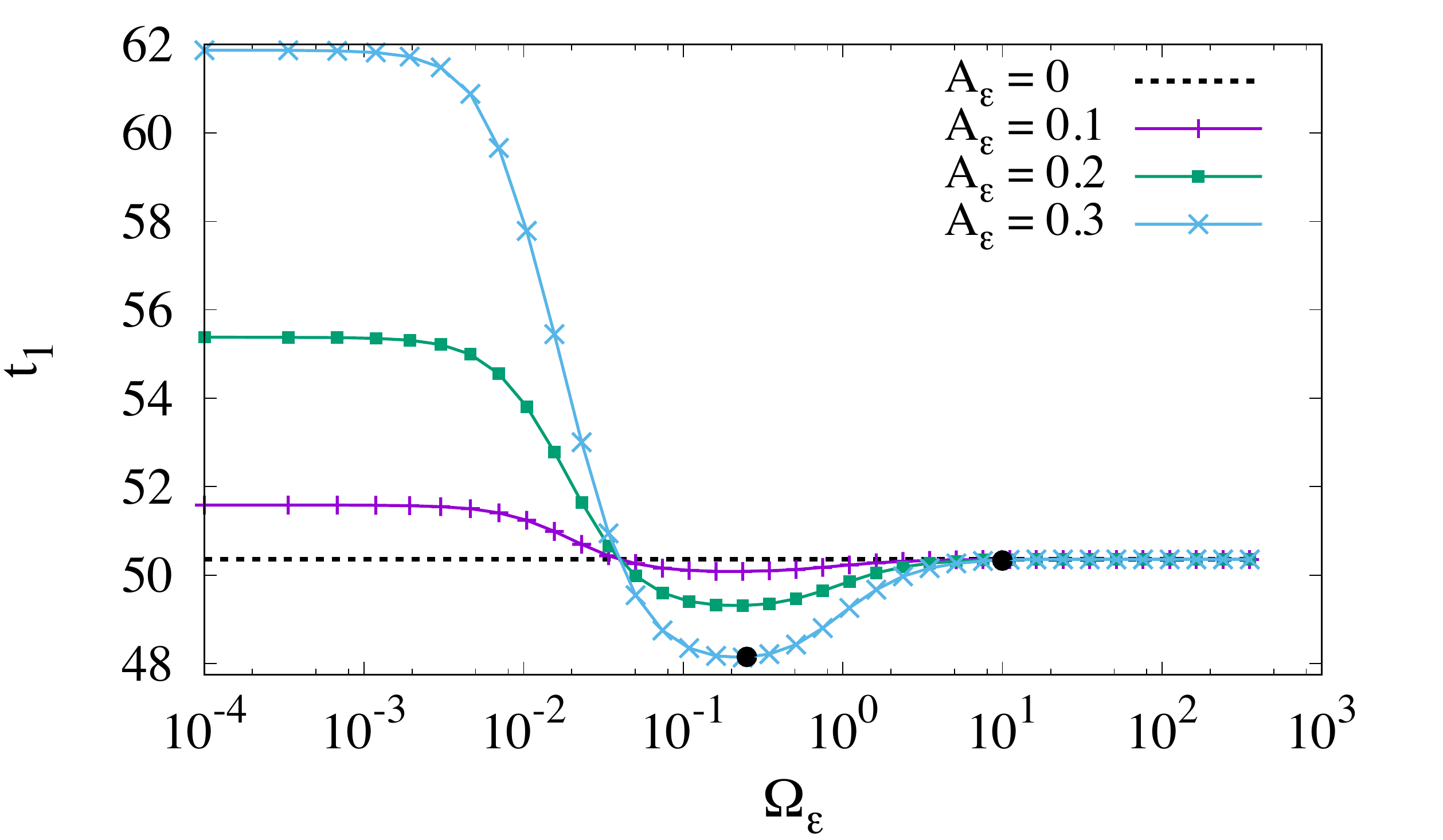}
\caption{\footnotesize{(Color online) Mean first passage time $t_1$ (averaged over initial driving phase $\phi$) {\it vs.} angular frequency $\Omega_{\epsilon}$ for a periodic driving of  the bias energy and  different driving strengths  $A_{\epsilon}$. Thee horizontal line marks again   the static case. The full (black)  circles highlight the values assumed by $t_1$ at  the two angular frequency values chosen for $\Omega_{\epsilon}$ in plotting the MFPT data in Fig.~\ref{fig13} below. Parameters $\alpha$, $T$, and $\omega_c$ are as in Fig.~\ref{fig2}.}}
\label{fig11}
\end{center}
\end{figure}
The  MFPT results versus angular driving frequency in Fig.~\ref{fig11} overall share the same features with those for noise-driven and periodically driven tunneling matrix element shown in Figs.~\ref{fig3} and~\ref{fig7}, respectively.
\subsection{Periodically oscillating bias and two-state fluctuating tunneling matrix element: Numerical treatment}
\label{driven-bias-and-Delta}
\indent In this subsection we consider the combined action of   dichotomous noise   and a periodic  driving. Specifically, we consider the MFPT $t_1$ as a function of the noise switching rate  $\nu$ of two-state noise on the tunneling matrix element, detailed by Eq.~(\ref{Delta_noise}), while simultaneously rocking periodically the  bias at angular frequency $\Omega_{\epsilon}$, according to Eq.~(\ref{epsilon-t}).\\
\begin{figure}[ht!]
\begin{center}
\includegraphics[width=0.5\textwidth,angle=0]{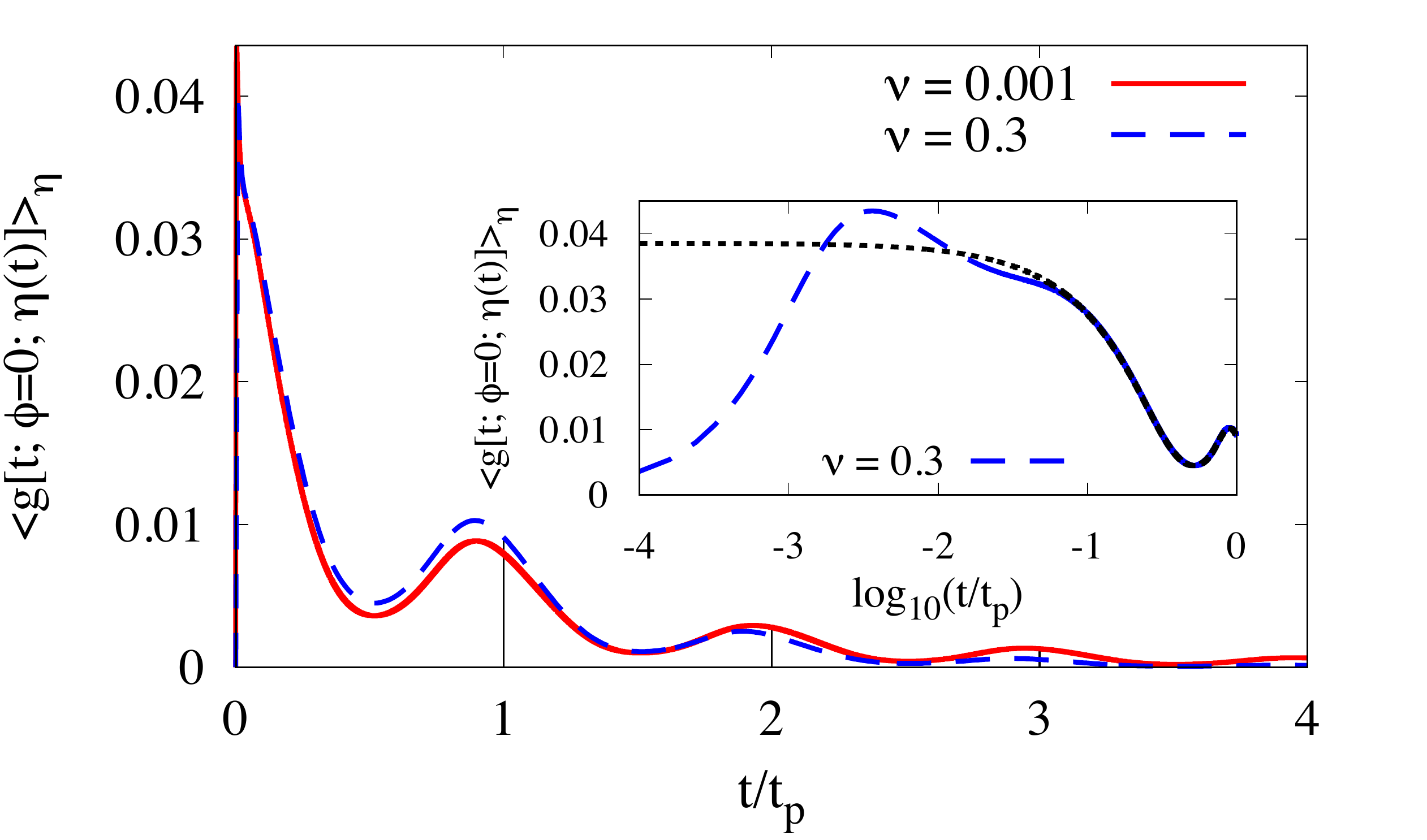}
\caption{\footnotesize{(Color online) First passage time pdf for periodic  driving of the bias $\epsilon(t)=A_{\epsilon}\cos(\Omega_{\epsilon} t+\phi)$ of amplitude $A_{\epsilon}=0.3$, period  $t_p=2\pi/\Omega_{\epsilon}$ with  $\Omega_{\epsilon}=0.1$ and initial driving phase  $\phi=0$.  The two-state noise of amplitude strength  $\Delta=0.3$ acts on the tunneling matrix element with a corresponding switching rate  $\nu$.  Inset: $\nu=0.3$ curve up to one driving period in $\mathrm \log_{10}$ scale. Dotted line: Same quantity evaluated using the rates calculated according to Eq.~(\ref{W-pm-i}). Other parameters are as in Fig.~\ref{fig2}.}}
\label{fig12}
\end{center}
\end{figure}
\indent Fig.~\ref{fig12} depicts the dynamics of the FPT pdf $g(t;\phi)=\langle g[t;\phi;\eta(t)]\rangle_{\eta}$ for a fixed initial driving phase $\phi=0$ and for two values of the Poisson parameter of the telegraphic noise modulating  the tunneling matrix element. As in Figs.~\ref{fig6} and~\ref{fig10}, $g(t;\phi)$ is modulated due to the presence of the deterministic periodic driving. This time, however, the additional presence of two-state noise, plotted for the same two switching rate parameters $\nu$, as done in Fig.~\ref{fig2}, affects the average behavior;  it does, however, not  wash out the  multi-peak behavior imposed by the applied periodic forcing.\\
\begin{figure}[ht!]
\begin{center}
\includegraphics[width=0.5\textwidth,angle=0]{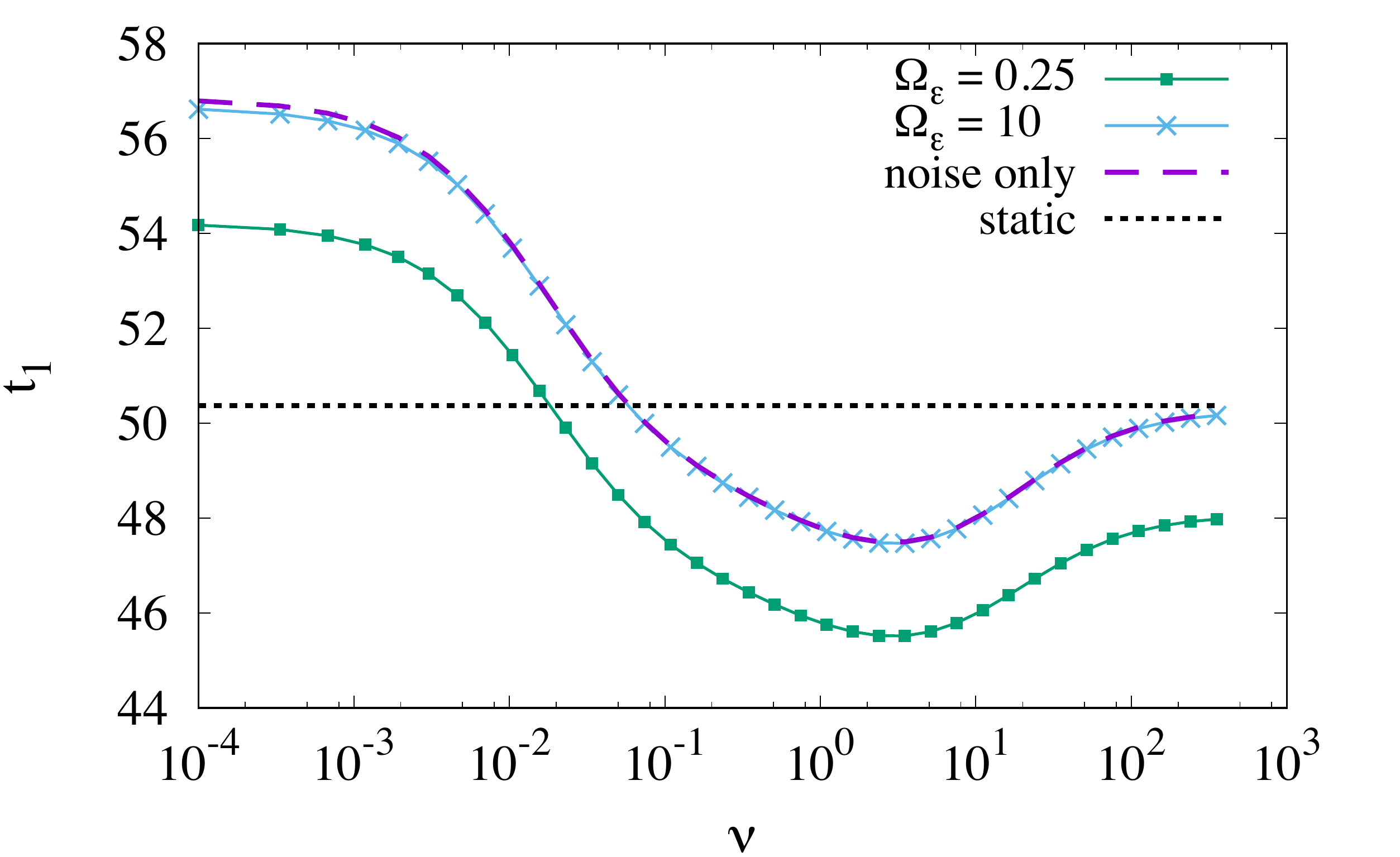}
\caption{\footnotesize{(Color online) Mean first passage time $t_1$ (averaged over initial phase $\phi$) versus  the two-state noise switching rate $\nu$ acting  on the tunneling matrix element in the presence of a simultaneous periodic driving of the bias $\epsilon(t)$.  The bias amplitude is held at $A_{\epsilon} = 0.3$ while the two  chosen angular frequencies values for $\Omega_{\epsilon}$ are indicated in the figure. In addition, a comparison is made with  the case  in which the deterministic drive for the bias  is switched off; i.e. dichotomous noise is solely modulating  the tunneling matrix element.  The dotted line marks the static case.  The amplitude for the modulation of the tunneling matrix amplitude is set at  $\Delta=0.2$. Other parameters are as in Fig.~\ref{fig2}.}}
\label{fig13}
\end{center}
\end{figure}
\indent In the setting considered here, the MFPT $t_1$ is obtained by solving Eq.~(\ref{eqPLfluct}) with transition rates given by Eq.~(\ref{W-pm-i}).
Our findings are shown in Fig.~\ref{fig13} for two angular driving frequencies of the bias. These two chosen values for $\Omega_{\epsilon}$ are  marked by filled circles in Fig.~\ref{fig11}; see curve of MFPT at $A_{\epsilon} = 0.3$. A further comparison is  made  with the noise-only case, i.e. with the  periodic driving being  switched off. Also here the average over the initial driving phase $\phi$ detailed by Eq.~(\ref{g-driving-noise}) is performed by uniformly sampling the interval $[0,2\pi)$ at $40$ points.\\
\indent  While the overall behavior of $t_1(\nu)$ exhibits the same features observed as in subsection IV.A, the role of introducing a periodically driven bias with frequency $\Omega_{\epsilon}$ consists in shifting  downwards to smaller values the curves of the MFPT versus the switching rate $\nu$. Specifically, $t_1$ assumes systematically lower values with a bias $\epsilon(t)$ periodically driven at $\Omega_{\epsilon}=0.25$.  At this angular driving frequency, $t_1(\nu)$ converges in the limit $\nu \rightarrow \infty$ to the  value highlighted by the full circle located at the minimum of $t_1$ in Fig.~\ref{fig11}, as to be expected. Likewise, for the case of a large  angular driving frequency, i.e., $\Omega_{\epsilon}=10$, the line $t_1(\nu)$ virtually coincides with the  analytical result obtained with $\epsilon(t)=0$ and dichotomous noise on the tunneling matrix element. This is due to the fact that, for such a large deterministic driving frequency, one approaches the situation discussed in Fig.~\ref{fig3} (green line) for $\Delta=0.2$.
\section{Conclusions}
\label{conclusions}
With this work we studied, by the use of analytical and numerical means, the phenomenon of resonant activation, occurring for a dissipative two-state quantum system (spin-boson system) which is modulated by  periodic deterministic driving and/or via telegraphic two-state noise. At strong system-bath coupling the quantum dynamics proceeds incoherently   so that an effective classical description in terms of a master equation with incoherent quantum rates becomes feasible. This in turn allows for studying the detailed first passage time statistics when  starting out at one of the two metastable states, with  absorption occurring at the neighboring state.\\
\indent Here we studied the complete first passage time probability density for general time-dependent driving of the two energy parameters characterizing the  spin-boson system. Specific driving mechanisms involve a modulation in terms of a stationary two-state process with exponentially correlated noise or also an external  deterministic periodic driving of those parameters, including combinations of both driving mechanisms.  In contrast to the case of stationary noise driving, the passage time dynamics for deterministic driving is cumbersome as it involves explicit time-dependent transition rates with corresponding time-dependent boundary conditions for reflection and absorption. Particularly, the role of periodic driving results in a decaying  first passage time probability densities which exhibits multiple peaks. These peaks reflect an  initial phase-dependent quantum synchronization feature \cite{Schindler2004,Schindler2005,Goychuk2006}. This latter feature is absent when the transition rates are time-independent (stationary noise driving), resulting now in a monotonic decay of the first passage time pdf.\\
\indent This first passage time pdf  allows for the evaluation of all its moments. Of particular interest is its first mean, the MFPT. This quantity displays the typical signatures of resonant activation, i.e.  the existence of an intermediate modulation regime where the MFPT undrerruns the values assumed in the opposite limits of  adiabatic slow driving and high frequency modulation. In the limit of very high frequency modulation one approaches the non-driven MFPT value.\\
\indent Our findings for various modulation settings corroborate  the {\it universal}  behavior \cite{Pechukas1994} found for classical over-the-barrier resonant activation,  where (i) at low frequencies the MFPT is dominated by the adiabatic configuration, with the largest possible passage time ruling the overall escape, while (ii) for high frequency modulations the MFPT is governed by the value of the time-averaged energy profile -- yielding typically the static MFPT value --; (iii) for modulations at intermediate time-scales (of the order of the system dynamics time scale) the regime with minimal MFPT values emerges (resonant activation regime) where the MFPT underruns both limits (i) and (ii).  The wide parameter region for the  quantum tunneling rate in the modulated TSS allows one to engineer the regime of resonant activation towards either smaller or also -- more interestingly -- much wider modulation regimes. This feature becomes apparent by supplementing the information contained in Figs.~\ref{fig3} and~\ref{fig7} with those of Figs.~\ref{fig5} and~\ref{fig8}.\\
\indent A further interesting feature we detected with this study is the approximate, although nearly exact, crossing behavior (as demonstrated analytically and validated numerically in Sec. IV.A) of the nonadiabatic  MFPT entering the resonant activation regime at some critical frequency $\nu^*$, being only weakly dependent on the driving amplitude.\\
\indent The experimental implementation of an absorbing state may not always be straightforward. In such cases, the pdf of residence times provided by Eq.~(\ref{RTpdf}), or also the interspike pdf, i.e., the pdf of time intervals between transitions, are experimentally more readily available for analysis~\cite{Gammaitoni1998,Schindler2004,Talkner2005}, as compared to  the FPT pdf. These additional pdfs can be related to the FPT pdf via averages involving the asymptotic entrance time pdf for state $L$~\cite{Talkner2003,Schindler2004,Schindler2005,Talkner2005}.\\
\indent Candidates for experimentally establishing the resonant activation regime in the presence of dissipative tunneling are quantum dot systems, with the setup realized for the recent experiment reported in Ref.~\cite{Wagner2017}. These systems possess two key features: First, the possibility of real-time detecting the tunneling of individual charges \emph{in} and \emph{out of} the dot (source$\rightarrow$dot and dot$\rightarrow$drain). Second, highly controllable tunneling rates ensuring that, for suitable configurations, the backtunneling to the dot is negligible due to Coulomb repulsion, which corresponds to a zero backward rate in our model. Moreover, the controllability of the tunneling rates allows in principle for implementing modulation settings like those discussed here.\\
\indent In a different experiment~\cite{Han2001}, a time-resolved detection of tunneling out of a metastable potential well, which traps the zero voltage state of a superconducting Josephson tunnel junction, is performed. There, a bi-exponential survival probability in the well, signature of the so-called two-level decay-tunneling process, is found. This feature is due to an internal decay process dependent on temperature, dissipation, and the internal level spacing set by the (tunable) barrier. A similar behavior is  found  for our model in the noise-only case, where, being no inside-well structures present,  the double-exponential decay is determined by the noise on the tunneling element and reduces to a single exponential in the limit of zero noise amplitude, as can be seen by inspection of  Eq.~(\ref{solution}).\\
\indent From the theoretical side, the present approach can be readily generalized to situations with many intermediate quantum states (overdamped tight-binding systems). However open challenges remain. A particularly difficult objective to be addressed in the future is its extension to the regime of  quantum coherence; i.e. to the case in which modulations act on weakly damped quantum systems. In this latter regime the very concept of a  (quasi-) classical MFPT analysis is doomed to fail.
\section{Acknowledgements}
The authors would like to thank Prof. Dr. P. Talkner for helpful discussions. P.H. acknowledges support by the Deutsche Forschungsgemeinschaft (DFG) via grant HA1517/35-1 and by the Singapore Ministry of Education and the National Research Foundation of Singapore. B. S. and D. V. acknowledge support by Ministry of Education, University and Research of Italian Government.
\appendix
\section{Derivation of the noise-averaged ME}
\label{derivation}
Using a reasoning put forward in  Ref.~\cite{Goychuk2005}, a dichotomous noise allows for an exact averaging of the dynamics of the population difference $P(t)$, which results in a set of equations where $\langle P(t)\rangle_{\eta}$ is coupled to the correlation expression $\langle P(t)\eta(t)\rangle_{\eta}$.\\
\indent  Along the same lines we derive Eq.~(\ref{eqPLfluct})  via an averaging of the equation for $P_L(t)$, with $R$ being an absorbing state,  over the noise realizations $\eta(t)$ of the dichotomous two-state  process
\begin{eqnarray}\label{noise}
\Delta(t)=\Delta_0+\Delta\eta(t)
\end{eqnarray}
detailed in Sec.~\ref{noise-av-ME}. We start out from
\begin{eqnarray}\label{eq}
\dot{P}_L(t)=W^+(t)P_L(t)
\end{eqnarray}
where
\begin{eqnarray}\label{W-p}
W^+(t)&=&\frac{\Delta(t)}{2}\int_0^{\infty}d\tau\; \Delta(t-\tau) e^{-Q'(\tau)}\nonumber\\
&&\qquad\times\cos[Q''(\tau)-\zeta(t,t-\tau)].
\end{eqnarray}
Substituting Eq.~(\ref{noise}) into Eq.~(\ref{W-p}) and performing the average over the noise we obtain
\begin{eqnarray}\label{eq-av}
\langle \dot{P}_L(t) \rangle_{\eta}&=&-\langle W^+(t)P_L(t) \rangle_{\eta}\nonumber\\
&=&-W^+_0(t)\langle P_L(t) \rangle_{\eta}-W^+_1(t) y(t),
\end{eqnarray}
where $y(t)=\langle \eta(t)P_L(t) \rangle_{\eta}$ and the rates $W^+_{0/1}$ are given in Eq.~(\ref{W-pm-i}). In passing from first to second line of Eq.~(\ref{eq-av}) we made use of two results in Ref.~\cite{Bourret1973}. The first is
\begin{eqnarray}\label{result1}
\langle \eta(t)\eta(t_1)\varPhi[\eta(\;)] \rangle_{\eta}&=&\langle \eta(t)\eta(t_1)\rangle_{\eta}\langle \varPhi[\eta(\;)] \rangle_{\eta}\qquad
\end{eqnarray}
with $t\geq t_1$, where $\varPhi[\eta(\;)]$ is a functional of the dichotomous noise involving times $\leq t_1$. Choosing $\varPhi[\eta(\;)]\equiv \eta(t_1)P_L(t_1)$
and using the properties $\eta^2(t)=1$ and  $\langle \eta(t)\eta(t') \rangle_{\eta}=\exp(-\nu|t-t'|)$, we obtain one of the identities necessary to derive Eq.~(\ref{eq-av}), namely
\begin{eqnarray}\label{result1-a}
\langle \eta(t-\tau)P_L(t) \rangle_{\eta}&=&\langle \eta(t)\eta(t-\tau)\rangle_{\eta}\langle \eta(t)P_L(t)\rangle_{\eta} \qquad
\end{eqnarray}
($\tau\leq t$).
The second result in Ref.~\cite{Bourret1973} reads
\begin{eqnarray}\label{result2}
&\langle \varPhi[\eta(\;)] \eta(t)\eta(t_1)\chi[\eta(\;)] \rangle_{\eta}=\langle  \varPhi[\eta(\;)] \eta(t)\rangle_{\eta}\langle \eta(t_1)\chi[\eta(\;)] \rangle_{\eta}&\nonumber\\
&+\langle  \varPhi[\eta(\;)] \rangle_{\eta}
 \langle \eta(t)\eta(t_1)\langle\chi[\eta(\;)] \rangle_{\eta}&
\end{eqnarray}
with $t\geq t_1$, where $\varPhi[\eta(\;)]$ and $\chi[\eta(\;)]$ are two functionals of the dichotomous noise involving times $\geq t$ and $\leq t_1$, respectively. Taking $\varPhi[\eta(\;)]\equiv P_L(t)$ and $\chi[\eta(\;)]\equiv1$, and using the property $\langle \eta(t) \rangle_{\eta}=0$, we get the identity
\begin{eqnarray}\label{result2-a}
\langle \eta(t-\tau)\eta(t)P_L(t) \rangle_{\eta}&=&\langle \eta(t)\eta(t-\tau)\rangle_{\eta}\langle P_L(t)\rangle_{\eta}\qquad
\end{eqnarray}
with $\tau\leq t$. Eq.~(\ref{result2-a}) is used, along with Eq.~(\ref{result1-a}), to obtain Eq.~(\ref{eq-av}).\\
\indent Next, the equation for  $y(t)=\langle \eta(t)P_L(t) \rangle_{\eta}$ can be derived  analogously starting from the theorem~\cite{Hanggi1978,Shapiro1978,Hanggi1985} which states that
\begin{equation}\label{}
\frac{d}{dt}\langle \eta(t)P_L(t) \rangle_{\eta}=-\nu\langle \eta(t)P_L(t) \rangle_{\eta}+\langle \eta(t)\dot{P}_L(t) \rangle_{\eta}.\qquad
\end{equation}
\indent  Using Eq.~(\ref{eq}) for $\dot{P}_L(t)$ on the rhs, calculating the noise averages by means of  Eqs.~(\ref{result1-a}) and~(\ref{result2-a}), and observing again that  $\eta^2(t)=1$, we find the following equation for $\dot{y}(t)$
\begin{eqnarray}\label{eq-y}
\dot{y}(t)=-W^+_1(t)\langle P_L(t) \rangle_{\eta}-[\nu + W^+_0(t) ]y(t).\quad
\end{eqnarray}
%
%\bibliography{bibliography}

\end{document}